\documentclass[fleqn,usenatbib]{mnras}

\usepackage{newtxtext,newtxmath}

\usepackage[T1]{fontenc}

\DeclareRobustCommand{\VAN}[3]{#2}
\let\VANthebibliography\thebibliography
\def\thebibliography{\DeclareRobustCommand{\VAN}[3]{##3}\VANthebibliography}

\usepackage{graphicx}	
\usepackage{amsmath}	
\usepackage{hyperref}
\usepackage{booktabs}
\usepackage{natbib}

\usepackage{color}
\usepackage{xcolor}
\definecolor{byzantine}{rgb}{0.74, 0.2, 0.64}

\title[TOI-561 architecture]{Architecture of TOI-561 planetary system \thanks{CHEOPS Program MR.Improve-CHESS ID: 0031}}

\author[G. Piotto et al.]{
G.~Piotto,$^{1, 2}$\,$^{\href{https://orcid.org/0000-0002-9937-6387}{\protect\includegraphics[height=0.19cm]{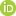}}}$
T.~Zingales,$^{1, 2}$\,$^{\href{https://orcid.org/0000-0001-6880-5356}{\protect\includegraphics[height=0.19cm]{orcid.jpg}}}$
L.~Borsato,$^{2}$\,$^{\href{https://orcid.org/0000-0003-0066-9268}{\protect\includegraphics[height=0.19cm]{orcid.jpg}}}$
J.~A.~Egger,$^{3}$\,$^{\href{https://orcid.org/0000-0003-1628-4231}{\protect\includegraphics[height=0.19cm]{orcid.jpg}}}$
A.~C.~M.~Correia,$^{4}$\,$^{\href{https://orcid.org/0000-0002-8946-8579}{\protect\includegraphics[height=0.19cm]{orcid.jpg}}}$
A.~E.~Simon,$^{3, 5}$\,$^{\href{https://orcid.org/0000-0001-9773-2600}{\protect\includegraphics[height=0.19cm]{orcid.jpg}}}$
\newauthor
H.-G.~Florén,$^{6}$
S.~G.~Sousa,$^{7}$\,$^{\href{https://orcid.org/0000-0001-9047-2965}{\protect\includegraphics[height=0.19cm]{orcid.jpg}}}$
P.~F.~L.~Maxted,$^{8}$\,$^{\href{https://orcid.org/0000-0003-3794-1317}{\protect\includegraphics[height=0.19cm]{orcid.jpg}}}$
D.~Nardiello,$^{1, 2}$\,$^{\href{https://orcid.org/0000-0003-1149-3659}{\includegraphics[scale=0.5]{orcid.jpg}}}$
L.~Malavolta,$^{1, 2}$\,$^{\href{https://orcid.org/0000-0002-6492-2085}{\includegraphics[scale=0.5]{orcid.jpg}}}$
T.~G.~Wilson,$^{9}$\,$^{\href{https://orcid.org/0000-0001-8749-1962}{\protect\includegraphics[height=0.19cm]{orcid.jpg}}}$
\newauthor
Y.~Alibert,$^{5, 3}$\,$^{\href{https://orcid.org/0000-0002-4644-8818}{\protect\includegraphics[height=0.19cm]{orcid.jpg}}}$
V.~Adibekyan,$^{7}$\,$^{\href{https://orcid.org/0000-0002-0601-6199}{\protect\includegraphics[height=0.19cm]{orcid.jpg}}}$
A.~Bonfanti,$^{10}$\,$^{\href{https://orcid.org/0000-0002-1916-5935}{\protect\includegraphics[height=0.19cm]{orcid.jpg}}}$
R.~Luque,$^{11}$
N.~C.~Santos,$^{7, 12}$\,$^{\href{https://orcid.org/0000-0003-4422-2919}{\protect\includegraphics[height=0.19cm]{orcid.jpg}}}$
M.~J.~Hooton,$^{13}$\,$^{\href{https://orcid.org/0000-0003-0030-332X}{\protect\includegraphics[height=0.19cm]{orcid.jpg}}}$
\newauthor
L.~Fossati,$^{10}$\,$^{\href{https://orcid.org/0000-0003-4426-9530}{\protect\includegraphics[height=0.19cm]{orcid.jpg}}}$
A.~M.~S.~Smith,$^{15}$\,$^{\href{https://orcid.org/0000-0002-2386-4341}{\protect\includegraphics[height=0.19cm]{orcid.jpg}}}$
S.~Salmon,$^{16}$\,$^{\href{https://orcid.org/0000-0002-1714-3513}{\protect\includegraphics[height=0.19cm]{orcid.jpg}}}$
G.~Lacedelli,$^{17}$\,$^{\href{https://orcid.org/0000-0002-4197-7374}{\protect\includegraphics[height=0.19cm]{orcid.jpg}}}$
R.~Alonso,$^{17, 19}$\,$^{\href{https://orcid.org/0000-0001-8462-8126}{\protect\includegraphics[height=0.19cm]{orcid.jpg}}}$
\newauthor
T.~Bárczy,$^{20}$\,$^{\href{https://orcid.org/0000-0002-7822-4413}{\protect\includegraphics[height=0.19cm]{orcid.jpg}}}$
D.~Barrado~Navascues,$^{21}$\,$^{\href{https://orcid.org/0000-0002-5971-9242}{\protect\includegraphics[height=0.19cm]{orcid.jpg}}}$
S.~C.~C.~Barros,$^{7, 12}$\,$^{\href{https://orcid.org/0000-0003-2434-3625}{\protect\includegraphics[height=0.19cm]{orcid.jpg}}}$
W.~Baumjohann,$^{10}$\,$^{\href{https://orcid.org/0000-0001-6271-0110}{\protect\includegraphics[height=0.19cm]{orcid.jpg}}}$
T.~Beck,$^{3}$
\newauthor
W.~Benz,$^{3, 5}$\,$^{\href{https://orcid.org/0000-0001-7896-6479}{\protect\includegraphics[height=0.19cm]{orcid.jpg}}}$
N.~Billot,$^{16}$\,$^{\href{https://orcid.org/0000-0003-3429-3836}{\protect\includegraphics[height=0.19cm]{orcid.jpg}}}$
A.~Brandeker,$^{6}$\,$^{\href{https://orcid.org/0000-0002-7201-7536}{\protect\includegraphics[height=0.19cm]{orcid.jpg}}}$
C.~Broeg,$^{3, 5}$\,$^{\href{https://orcid.org/0000-0001-5132-2614}{\protect\includegraphics[height=0.19cm]{orcid.jpg}}}$
M.~Buder$^{15}$,
A.~Collier~Cameron,$^{22}$\,$^{\href{https://orcid.org/0000-0002-8863-7828}{\protect\includegraphics[height=0.19cm]{orcid.jpg}}}$
\newauthor
Sz.~Csizmadia,$^{15}$\,$^{\href{https://orcid.org/0000-0001-6803-9698}{\protect\includegraphics[height=0.19cm]{orcid.jpg}}}$
P.~E.~Cubillos,$^{10, 23}$
M.~B.~Davies,$^{24}$\,$^{\href{https://orcid.org/0000-0001-6080-1190}{\protect\includegraphics[height=0.19cm]{orcid.jpg}}}$
M.~Deleuil,$^{25}$\,$^{\href{https://orcid.org/0000-0001-6036-0225}{\protect\includegraphics[height=0.19cm]{orcid.jpg}}}$
A.~Deline,$^{16}$
\newauthor
O.~D.~S.~Demangeon,$^{7, 12}$\,$^{\href{https://orcid.org/0000-0001-7918-0355}{\protect\includegraphics[height=0.19cm]{orcid.jpg}}}$
B.-O.~Demory,$^{5, 3}$\,$^{\href{https://orcid.org/0000-0002-9355-5165}{\protect\includegraphics[height=0.19cm]{orcid.jpg}}}$
A.~Derekas,$^{26}$
B.~Edwards,$^{27}$
D.~Ehrenreich,$^{16, 28}$\,$^{\href{https://orcid.org/0000-0001-9704-5405}{\protect\includegraphics[height=0.19cm]{orcid.jpg}}}$
A.~Erikson,$^{15}$
\newauthor
A.~Fortier,$^{3, 5}$\,$^{\href{https://orcid.org/0000-0001-8450-3374}{\protect\includegraphics[height=0.19cm]{orcid.jpg}}}$
M.~Fridlund,$^{29, 30}$\,$^{\href{https://orcid.org/0000-0002-0855-8426}{\protect\includegraphics[height=0.19cm]{orcid.jpg}}}$
D.~Gandolfi,$^{31}$\,$^{\href{https://orcid.org/0000-0001-8627-9628}{\protect\includegraphics[height=0.19cm]{orcid.jpg}}}$
K.~Gazeas,$^{32}$
M.~Gillon,$^{33}$\,$^{\href{https://orcid.org/0000-0003-1462-7739}{\protect\includegraphics[height=0.19cm]{orcid.jpg}}}$
M.~Güdel,$^{34}$
\newauthor
M.~N.~Günther,$^{35}$\,$^{\href{https://orcid.org/0000-0002-3164-9086}{\protect\includegraphics[height=0.19cm]{orcid.jpg}}}$
A.~Heitzmann,$^{16}$\,$^{\href{https://orcid.org/0000-0002-8091-7526}{\protect\includegraphics[height=0.19cm]{orcid.jpg}}}$
Ch.~Helling,$^{10, 36}$
K.~G.~Isaak,$^{35}$\,$^{\href{https://orcid.org/0000-0001-8585-1717}{\protect\includegraphics[height=0.19cm]{orcid.jpg}}}$
L.~L.~Kiss,$^{37, 38}$
J.~Korth,$^{39}$\,$^{\href{https://orcid.org/0000-0002-0076-6239}{\protect\includegraphics[height=0.19cm]{orcid.jpg}}}$
\newauthor
K.~W.~F.~Lam,$^{15}$\,$^{\href{https://orcid.org/0000-0002-9910-6088}{\protect\includegraphics[height=0.19cm]{orcid.jpg}}}$
J.~Laskar,$^{40}$\,$^{\href{https://orcid.org/0000-0003-2634-789X}{\protect\includegraphics[height=0.19cm]{orcid.jpg}}}$
A.~Lecavelier~des~Etangs,$^{41}$\,$^{\href{https://orcid.org/0000-0002-5637-5253}{\protect\includegraphics[height=0.19cm]{orcid.jpg}}}$
M.~Lendl,$^{16}$\,$^{\href{https://orcid.org/0000-0001-9699-1459}{\protect\includegraphics[height=0.19cm]{orcid.jpg}}}$
P.~Leonardi,$^{42, 1}$
D.~Magrin,$^{2}$\,$^{\href{https://orcid.org/0000-0003-0312-313X}{\protect\includegraphics[height=0.19cm]{orcid.jpg}}}$
\newauthor
G.~Mantovan$^{1, 2}$\,$^{\href{https://orcid.org/0000-0002-6871-6131}{\protect\includegraphics[height=0.19cm]{orcid.jpg}}}$
C.~Mordasini,$^{3, 5}$
V.~Nascimbeni,$^{2}$\,$^{\href{https://orcid.org/0000-0001-9770-1214}{\protect\includegraphics[height=0.19cm]{orcid.jpg}}}$
G.~Olofsson,$^{6}$\,$^{\href{https://orcid.org/0000-0003-3747-7120}{\protect\includegraphics[height=0.19cm]{orcid.jpg}}}$
R.~Ottensamer,$^{55}$
I.~Pagano,$^{44}$\,$^{\href{https://orcid.org/0000-0001-9573-4928}{\protect\includegraphics[height=0.19cm]{orcid.jpg}}}$
\newauthor
E.~Pallé,$^{17, 19}$\,$^{\href{https://orcid.org/0000-0003-0987-1593}{\protect\includegraphics[height=0.19cm]{orcid.jpg}}}$
G.~Peter,$^{45}$\,$^{\href{https://orcid.org/0000-0001-6101-2513}{\protect\includegraphics[height=0.19cm]{orcid.jpg}}}$
R.~Ottensamer,$^{54}$
D.~Pollacco,$^{9}$
D.~Queloz,$^{46, 13}$\,$^{\href{https://orcid.org/0000-0002-3012-0316}{\protect\includegraphics[height=0.19cm]{orcid.jpg}}}$
R.~Ragazzoni,$^{2, 1}$\,$^{\href{https://orcid.org/0000-0002-7697-5555}{\protect\includegraphics[height=0.19cm]{orcid.jpg}}}$
N.~Rando,$^{35}$
\newauthor
F.~Ratti,${35}$
H.~Rauer,$^{15, 47}$\,$^{\href{https://orcid.org/0000-0002-6510-1828}{\protect\includegraphics[height=0.19cm]{orcid.jpg}}}$
I.~Ribas,$^{48, 49}$\,$^{\href{https://orcid.org/0000-0002-6689-0312}{\protect\includegraphics[height=0.19cm]{orcid.jpg}}}$
G.~Scandariato,$^{44}$\,$^{\href{https://orcid.org/0000-0003-2029-0626}{\protect\includegraphics[height=0.19cm]{orcid.jpg}}}$
D.~Ségransan,$^{16}$\,$^{\href{https://orcid.org/0000-0003-2355-8034}{\protect\includegraphics[height=0.19cm]{orcid.jpg}}}$
D.~Sicilia,$^{53}$\,$^{\href{https://orcid.org/0000-0001-7851-5168}{\protect\includegraphics[height=0.19cm]{orcid.jpg}}}$
\newauthor
M.~Stalport,$^{50, 33}$
S.~Sulis,$^{25}$\,$^{\href{https://orcid.org/0000-0001-8783-526X}{\protect\includegraphics[height=0.19cm]{orcid.jpg}}}$
Gy.~M.~Szabó,$^{26, 51}$\,$^{\href{https://orcid.org/0000-0002-0606-7930}{\protect\includegraphics[height=0.19cm]{orcid.jpg}}}$
S.~Udry,$^{16}$\,$^{\href{https://orcid.org/0000-0001-7576-6236}{\protect\includegraphics[height=0.19cm]{orcid.jpg}}}$
S.~Ulmer-Moll,$^{10, 5, 50}$\,$^{\href{https://orcid.org/0000-0003-2417-7006}{\protect\includegraphics[height=0.19cm]{orcid.jpg}}}$
\newauthor
V.~Van~Grootel,$^{50}$\,$^{\href{https://orcid.org/0000-0003-2144-4316}{\protect\includegraphics[height=0.19cm]{orcid.jpg}}}$
J.~Venturini,$^{16}$\,$^{\href{https://orcid.org/0000-0001-9527-2903}{\protect\includegraphics[height=0.19cm]{orcid.jpg}}}$
E.~Villaver,$^{17, 19}$
N.~A.~Walton,$^{52}$\,$^{\href{https://orcid.org/0000-0003-3983-8778}{\protect\includegraphics[height=0.19cm]{orcid.jpg}}}$ 
J.~N.~Winn,$^{54}$\,$^{\href{https://orcid.org/0000-0002-4265-047X}{\protect\includegraphics[height=0.19cm]{orcid.jpg}}}$ and
S.~Wolf$^{16}$
\\
\\
(Affiliations can be found after the references)
}

\date{Accepted XXX. Received YYY; in original form ZZZ}

\pubyear{2022}

\begin{document}
\label{firstpage}
\pagerange{\pageref{firstpage}--\pageref{lastpage}}
\maketitle

\begin{abstract}
We present new observations from CHEOPS and TESS to clarify the architecture of the planetary system hosted by the old Galactic thick disk star TOI-561. Our global analysis, which also includes previously published photometric and radial velocity data, incontrovertibly proves that TOI-561 is hosting at least four transiting planets with periods of 0.44 days (TOI-561 b), 10.8 days (TOI-561 c), 25.7 days (TOI-561 d), and 77.1 days (TOI-561 e) and a fifth non-transiting candidate, TOI-561f with a period of 433 days. The precise characterisation of TOI-561’s orbital architecture is interesting since old and metal-poor thick disk stars are less likely to host ultra-short period Super-Earths like TOI-561\,b. The new period of planet -e is consistent with the value obtained using radial velocity alone and is now known to be $77.14399\pm0.00025$ days, thanks to the new CHEOPS and TESS transits. The new data allowed us to improve its radius  ($R_p = 2.517 \pm 0.045 R_{\oplus}$ from 5\% to 2\% precision)  and mass ($M_p = 12.4 \pm 1.4 M_{\oplus}$) estimates, implying a density of $\rho_p = 0.778 \pm 0.097 \rho_{\oplus}$. Thanks to recent TESS observations and the focused CHEOPS visit of the transit of TOI-561 e, a good candidate for exomoon searches, the planet's period is finally constrained, allowing us to predict transit times through 2030 with 20-minute accuracy. We present an updated version of the internal structure of the four transiting planets. We finally performed a detailed stability analysis, which confirmed the long-term stability of the outer planet TOI-561 f.
\end{abstract}



\clearpage

\section{Introduction}

The planetary system orbiting the old Galactic thick disc star TOI-561 is currently known to include an ultra-short period planet and at least three mini-Neptunes. There has been controversy in the literature over the correct architecture of this system. It represents a sort of unique situation in which the analysis of the same light curves and two radial velocity datasets led to two different solutions, in disagreement regarding the number of planets in the system, their masses, and even their periods \citep{lacedelli2021, wiess2021,lacedelli2022}.

More recently, \citet{lacedelli2022} detected a further planet candidate with a period of about 400 d based on HARPS-N radial velocity data. The existence of this fifth non-transiting planet needs to be confirmed with future observational surveys.


In this work, based on additional TESS \citep{ricker2015} and CHEOPS \citep{benz2021,fortier2024}, data, we confirm that TOI-561 is a system made of five planets, four of which have been detected by transits. Its host star is one of the oldest objects in our Galaxy ($t_* = 11.0_{-3.5}^{+2.8}$ Gyr, \citealp{lacedelli2022}) with a detected planetary system. TOI-561 is one of the 78\footnote{\url{https://exoplanetarchive.ipac.caltech.edu/}} planetary systems with at least four confirmed exoplanets and an ultra-short period planet (USP).
Refining the planetary structure and orbital configuration will open a window about the earlier planetary formation and evolution in our Galaxy. TOI-561 b stands out as a USP with extremely low density. TOI-561\,b seems to have lost most of its atmosphere and might host water. For further details on the structure of TOI-561 b see \citet{patel2023}. This planet was selected among the winning JWST Proposals in Cycle 2, ID. \#3860, PI: J. Teske.

Usually, young, slow, and metal-rich stars associated with the thin disk host averagely more exoplanets, especially close-in Super-Earths \citep{bashi2021}. It means that the TOI-561 exoplanetary system is very peculiar since it is unlikely for an old and metal-poor thick disk star to host a USP. Considering the properties of this system, the USP TOI-561\,b is likely composed of high mean molecular weight species \citep{brinkman2022}.

TOI-561\,f is the only non-transiting candidate exoplanet in this system. Radial velocity data suggest an orbital period of over 400 days. Correct knowledge of the properties of this fifth planet are very important to constrain the dynamics and formation mechanisms of planets around old stars like TOI-561. Its existence also raises interesting questions about the possibility of other transiting and non-transiting planets in the system that could be identified with long-term follow-up and more advanced methods.

The 4th transiting planet, TOI-561 e was first inferred from HARPS-N radial velocity. The ephemerides obtained from the RVs allowed \citet{lacedelli2021} to identify a corresponding transit of the $\sim 77 d$ period planet in TESS Sector 8. The values shown in \citet{lacedelli2021} helped us detect a new transit in Sector 45, which was indeed observed and visible in the TESS Image CAlibrator (TICA) Full Frame Images of this sector. With the use of the new ephemerides, we scheduled an observation of the TOI-561 system using the CHEOPS telescope. In this work, we present the refined ephemerides and the new planetary architecture of the TOI-561 system using the newest CHEOPS+TESS observations. We confirm the presence of the fourth transiting planet, clarifying the architecture of the system, and we present new ephemerides and analysis of the structure of the planets orbiting TOI-561. Because of its long period, TOI-561 e is one of the few good candidates for the search for exomoons and, therefore, suited for follow-up.

In Section \ref{sec:observations} we describe all the new observations. In Sections \ref{sec:analysis} we show the final analysis performed for this system, highlighting the improvement of the ephemerides of all targets (Sec \ref{sec:ephemerides}). In Section \ref{sec:stability} we perform a stability analysis of TOI-561 f's orbit, the outermost candidate exoplanet of this system. Finally, we show an improved internal structure model for all the transiting exoplanets (Section \ref{sec:internal}).

\section{Observations}\label{sec:observations}
We extracted the light curve of TOI-561 from the TICA\footnote{\url{https://archive.stsci.edu/hlsp/tica}} \citep{fausnaugh2021} images collected during Sector 45 (from 2021-Nov-06 to 2021-Dec-02). For the light curve extraction, we used the PATHOS pipeline developed by \citet{nardiello2019,nardiello2020b} adapted to TICA images. Light curves were corrected by applying to them Cotrending Basis Vectors obtained as in \citet{nardiello2020a}.

The TICA analysis of Sector 45 made it possible to refine the ephemerides and schedule a further transit of planet TOI-561 e with CHEOPS on February 7, 2022. We used four CHEOPS observations of the TOI-561 system that cover the transits of planets b, c, d, and e, from 2021-01-23 to 2022-02-08 (see Table \ref{tab:cheops_visits} for a chronology of the observations).

Additionally, we used TESS Full Frame Images (FFIs) of Sectors 46 (from 2021-12-02 to 2021-12-30) and 72 (from 2023-11-11 to 2023-12-07). We extracted the raw light curves from FFIs by using the pyPATHOS pipeline (see \citealt{nardiello2019,nardiello2021}) and we corrected them by applying the cotrending basis vectors obtained by \citet{nardiello2020}.

\begin{figure*}
    \centering
    \includegraphics[width=0.45\textwidth]{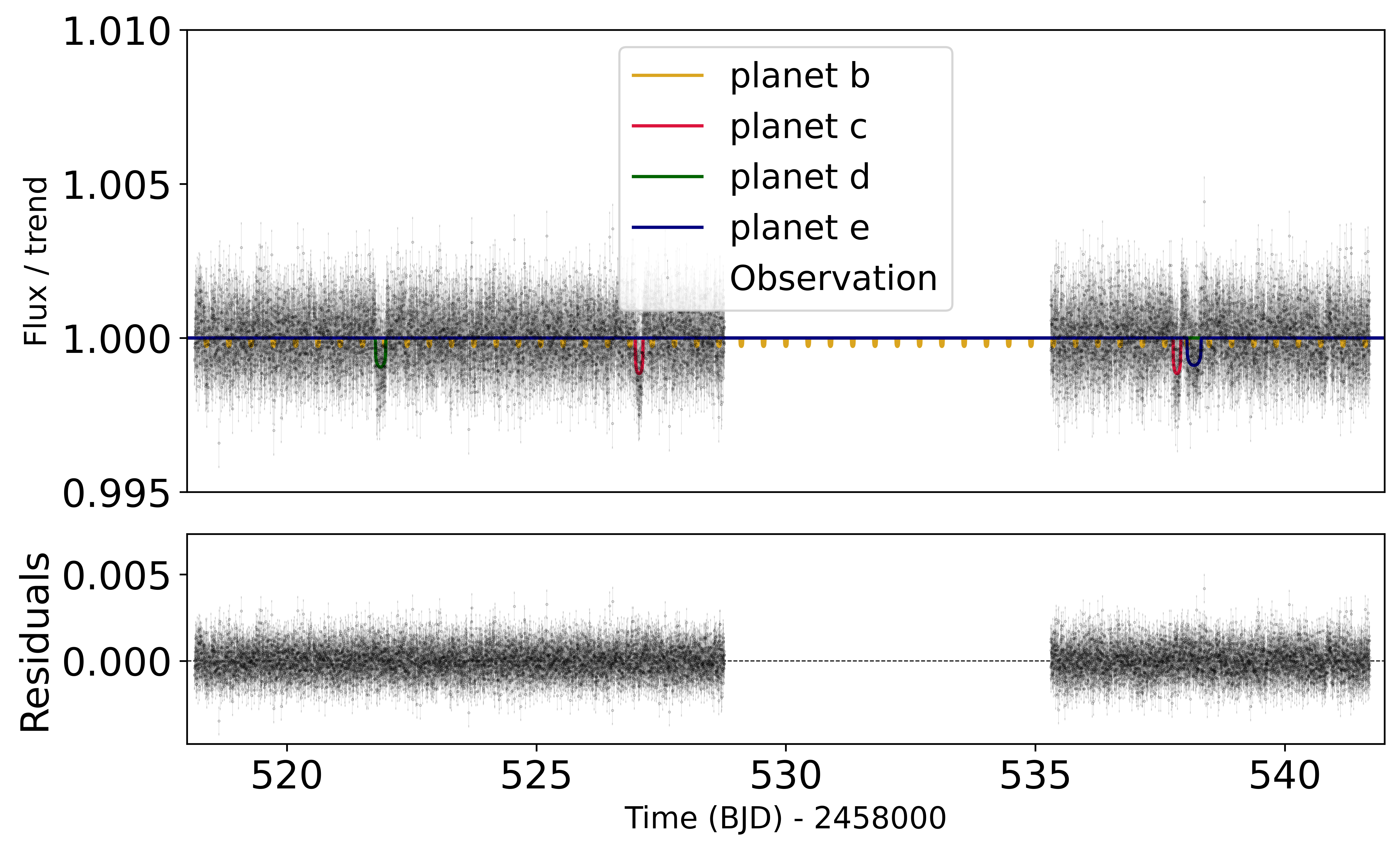}
    \includegraphics[width=0.45\textwidth]{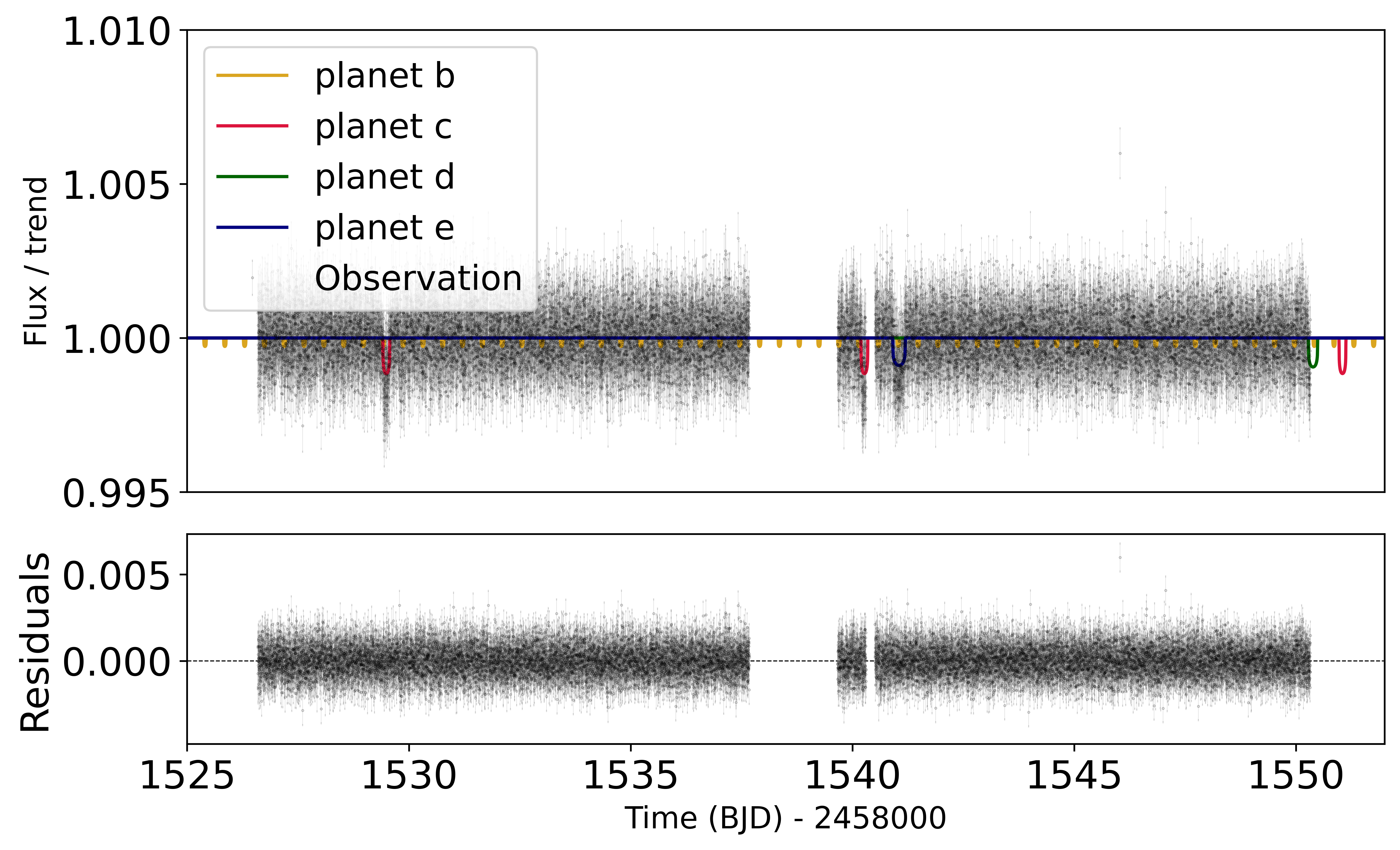}
    \caption{TOI-561 TESS observations. Different colors show the light curve of the different transiting planet models. \textbf{Left}: TESS Sector 8 with PyORBIT models. \textbf{Right}: TESS sector 45 with the \texttt{PyORBIT} models.}
    \label{fig:tess_sectors_obs}
\end{figure*}

\begin{table*}
\caption[]{CHEOPS observations of TOI-561 system. Each row provides the time window of the observation, with the starting and the end time of each visit, the CHEOPS data name, and the planets that fall into the specific visit.}
\label{tab:cheops_visits}
$$ 
\begin{tabular}{ccccccc}
\hline
\noalign{\smallskip}
Start [UTC] & End [UTC] & Duration [hrs] & Number of frames & Integration time [s] & File Key &  planet \\
\noalign{\smallskip}
\hline
\noalign{\smallskip}
2022-02-06 23:54 & 2022-02-08 06:03 & 30.16 & 1179 & 60 & CH\_PR100031\_TG049101\_V0200 & b,e \\
2021-04-12 23:52 & 2021-04-15 05:37 & 53.76 & 2057 & 60 & CH\_PR100031\_TG039301\_V0200 & b,d \\
2021-03-29 10:19 & 2021-03-29 14:44 & 4.42 & 212 & 60 &  CH\_PR100008\_TG000811\_V0200 & b,c \\
2021-01-23 15:29 & 2021-01-24 07:09 & 15.67 & 617 & 60 & CH\_PR100031\_TG037001\_V0200 & b,c \\
\noalign{\smallskip}
\hline
\end{tabular}
$$ 
\end{table*}

We used the published radial velocities of TOI-561, which have been obtained with HARPS-N at the Telescopio Nazionale Galileo \citep{cosentino2014}, and HIRES at Keck \citep{vogt1993}, for a total of $122$ and $61$ radial velocity measurements, respectively. We refer to \citet{lacedelli2021, lacedelli2022, wiess2021} for more details on spectroscopic data.


 
\section{Photometric and RV analysis}\label{sec:analysis}
We used the light curves delivered by the CHEOPS Data Reduction Pipeline (DRP v13, \citet{hoyer2020}). The CHEOPS DRP automatically calibrates and corrects the data and delivers the photometric light curves obtained with different aperture sizes. In this work, we used the DEFAULT (R=25) aperture which has the best light curve in terms of dispersion.
CHEOPS observations have been detrended using \texttt{cheope}\footnote{\url{https://github.com/tiziano1590/cheops_analysis-package}}, a \texttt{python} tool which uses \texttt{pycheops} \citep{maxted2021} as a backend and optimized to extract the planetary signal from the CHEOPS data frames. \texttt{cheope} uses \texttt{lmfit} \citep{newville2014} to select the best detrending models and the \texttt{emcee} \citep{foreman2013} package for the Bayesian framework. All CHEOPS light curves from \citet{lacedelli2021}  and \citet{lacedelli2022}, including the one from the last CHEOPS visit of TOI-561 e, were reanalyzed with \texttt{cheope}.

We used the PDCSAP TESS light curves (Fig. \ref{fig:tess_sectors_obs}) from sectors 8,35,45, 46 and 72 and flattened using \texttt{wotan} \citep{hippke2019}, which allow to model the stellar flux with a Tukey's biweight estimator \citep{tukey1977}.

The stellar radius was computed using the method described in \citet{lacedelli2022}. They used an MCMC IRFM method that compares broadband photometry to stellar spectral catalogs using the results of their spectral analysis, such as metallicity, as priors. Thanks to Gaia contribution with improved parallaxes and distances we can now get down to 1\% (internal) stellar radius uncertainties. This can also be seen in independent works using the ARIADNE code\footnote{\url{https://arxiv.org/abs/2204.03769},}\footnote{\url{https://github.com/jvines/astroARIADNE)}}.

To obtain an updated architecture of the TOI-561 system,  we performed a simultaneous modeling of all available photometric data from  CHEOPS and TESS, and spectroscopic data from HARPS-N and HIRES. We used the code \texttt{PyORBIT} \citep{malavolta2016, malavolta2018} with the  \texttt{PyDE}+\texttt{emcee} sampling method \citep{parviainen2016, foreman2013}. To find the best starting point for the MCMC sampling, we used PyDE with 64000 generators on our datasets. Then, we ran 100 chains for 150000 steps, discarding the first 40000 and applying a thinning factor of 100. We used the same priors and boundaries as reported in \citet{lacedelli2022} unless otherwise specified in this section.


We assumed four transiting planets plus a fifth non-transiting Keplerian applying a uniform prior on the period between 2 and 1000 days. We used the same stellar parameters as described in \citet{lacedelli2022}. The limb darkening has been parametrized using the quadratic law, as described in \citet{kipping2013}, using Gaussian priors on the $u_1$, $u_2$ coefficients, computed for both CHEOPS and TESS passbands, adopting a bilinear interpolation of the limb darkening profile as defined in \citet{claret2017,claret2021}. 
The Radial Velocity data points in this analysis are the same used in \citet{lacedelli2022}, so we do not report any significant change in the radial velocity semi-amplitude and, then, on the masses of the planets.

To exclude that our detrending was affecting the results, we also performed an independent global analysis with the \texttt{juliet} package \citep{espinoza2019} including all the available datasets and fitting all the planetary models simultaneously. The results from \texttt{PyORBIT} and \texttt{juliet} were consistent with each other, indicating that our assumptions and detrending did not significantly affect the results of our analysis. 

\begin{figure}
    \centering
    \includegraphics[width=0.5\textwidth]{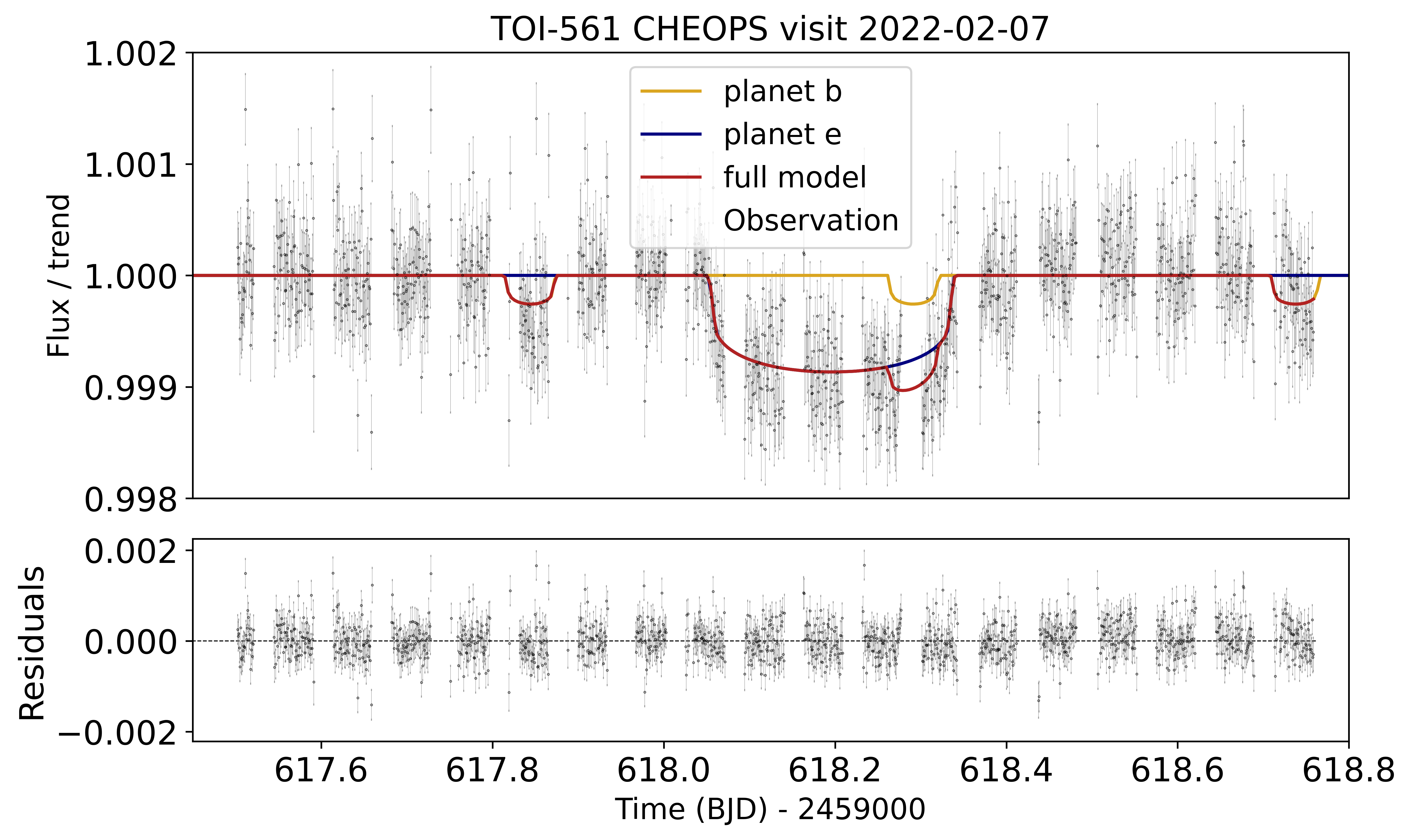}
    \caption{TOI-561 light curve from CHEOPS last visit of February 7, 2022.}
    \label{fig:cheops_last_visit}
\end{figure}

During the last visit with CHEOPS, also the ultra-short-period planet TOI-561~b was transiting in front of the star. The simultaneous fit of \texttt{PyORBIT} allows us to distinguish the two components of this transit (Fig \ref{fig:cheops_last_visit}). 

The new analysis, using latest new photometric data, improves our knowledge of the orbital parameters of the four transiting planets, i.e. TOI-561 b (P = $0.4465697 \pm 0.0000003$ days, $R_p = 1.397 \pm 0.027  R_{\oplus}$), c (P = $10.778838 \pm 0.000018$ days, $R_p = 2.865 \pm 0.041  R_{\oplus}$), d (P = $25.71268 \pm 0.00012$ days, $R_p = 2.615 \pm 0.059  R_{\oplus}$) and e (P = $77.14400 \pm 0.00027$ days, $R_p = 2.517 \pm 0.045 R_{\oplus}$) and refutes the existence of the planet at P = 16.3 days found in \citet{wiess2021}.

Our analysis resulted in an improvement of the radius of TOI-561 e with an error reduced to 2\%, thanks to the new TESS+CHEOPS observations. For comparison, in \citet{lacedelli2022} the precision on the radius was 5\%. \citet{lacedelli2022} have a precision on $P_e$ of about 6 hours, while in this work we reached a precision of 21.6 seconds.

\begin{figure}
    \centering
    \includegraphics[width=0.5\textwidth]{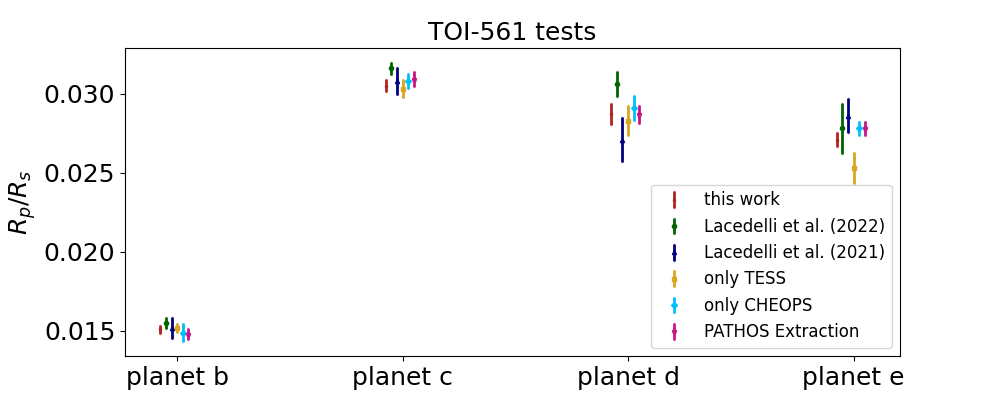}
    \caption{Comparison in the $R_p/R_*$ ratio for all the tests made in this letter. The figure illustrates the results obtained using only TESS data, only CHEOPS data, and from the independent analysis made by extracting TESS lightcurves with the PATHOS code.}
    \label{fig:test_comparison}
\end{figure}

In order to resolve the discrepancies appearing in the values of $R_p/R_*$ for TOI-561 c and d as reported by \citet{lacedelli2022}, we performed three different tests in an independent manner:
\begin{itemize}
    \item A global fit using only TESS data to check for inconsistencies arising from the CHEOPS dataset.
    \item A global fit using only CHEOPS data to check for inconsistencies arising from the TESS dataset.
    \item An independent analysis using the PATHOS code on all four TESS sectors.
\end{itemize}
The results of these tests, as presented in Fig \ref{fig:test_comparison}, confirm the final values of this work within 1$\sigma$.



Table \ref{tab:toi-561pars} lists the most updated parameters for the TOI-561 planetary system.

\begin{table*}
\caption[]{Parameters for the TOI-561 planetary system}
\label{tab:toi-561pars}
\centering
\resizebox{\textwidth}{!}{\begin{tabular}{cccccc}
    \toprule
    \toprule
    \multicolumn{6}{c}{Planetary parameters} \\
    \cmidrule{1-6}
     & TOI-561 b & TOI-561 c & TOI-561 d & TOI-561 e & 5$^{th}$ Keplerian \\
    \midrule
    $P [d]$  & $0.4465697\pm0.0000003$ & $10.778838\pm0.000018$ & $25.71268\pm0.00012$ & $77.14400\pm0.00027$ & $433^{+20}_{-18}$ \\ 
    $T_0$[TBJD]$^a$ & $2317.75002\pm0.00041$ & $2238.46284\pm0.00079$ & $2318.9711\pm0.0021$ & $1538.1803\pm0.0035$ & - \\ 
    $a/R_{*}$ & $2.683\pm0.029$ & $22.40\pm0.25$ & $40.00\pm0.44$ & $83.21\pm0.91$ & $262.9^{+8.5}_{-7.9}$ \\ 
    $a [AU]$ & $0.01064\pm0.00016$ & $0.0889\pm0.0013$ & $0.1587\pm0.0024$ & $0.3300\pm0.0050$ & $1.043\pm0.035$ \\ 
    $R_p/R_*$ & $0.01519\pm0.00028$ & $0.03115\pm0.00041$ & $0.02842\pm0.00063$ & $0.02736\pm0.00047$ & - \\ 
    $R_p [R_{\oplus}]$ & $1.397\pm0.027$ & $2.865\pm0.041$ & $2.615\pm0.059$ & $2.517\pm0.045$ & - \\ 
    $b$ & $0.14^{+0.096}_{-0.094}$ & $0.15^{+0.14}_{-0.10}$ & $0.36^{+0.13}_{-0.19}$ & $0.19^{+0.14}_{-0.13}$ & - \\ 
    $i$ [deg] & $87.0^{+2.0}_{-2.1}$ & $89.61^{+0.27}_{-0.33}$ & $89.51^{+0.25}_{-0.14}$ & $89.864^{+0.095}_{-0.094}$ & - \\ 
    $T_{14}$ [days] & $0.05452\pm0.00066$ & $0.1558^{+0.0022}_{-0.0042}$ & $0.197\pm0.012$ & $0.2974\pm^{+0.0057}_{-0.01}$ & - \\ 
    $e$ & 0 (fixed) & $0.023^{+0.034}_{-0.017}$ & $0.111^{+0.050}_{-0.039}$ & $0.074^{+0.044}_{-0.039}$ & $0.083^{+0.080}_{-0.058}$ \\ 
    $\omega$ [deg] & 90 (fixed)  & $219^{+87}_{-149}$ & $-131^{+17}_{-26}$ & $148^{+26}_{-39}$ & $-66^{+66}_{-83}$  \\ 
    $K [$m s$^{-1}]$ & $1.95\pm0.21$ & $1.98\pm0.21$ & $3.36\pm0.22$ & $2.16\pm0.23$ & $1.88\pm0.25$ \\ 
    $M_p [M_{\oplus}]^b$ & $2.02\pm0.23$ & $5.93\pm0.67$ & $13.33\pm0.98$ & $12.4\pm1.4$ & $19.1\pm2.7$ \\ 
    $\rho_p [\rho_{\oplus}]$ & $0.741\pm0.095$ & $0.252\pm0.030$ & $0.745\pm0.074$ & $0.778\pm0.097$ & - \\ 
    $\rho_p [g cm^{-3}]$ & $4.1\pm0.5$ & $1.3\pm0.2$ & $4.1\pm0.4$ & $4.3\pm0.5$ & - \\ 
    $T_{eq}$[K]$^c$ & $2319\pm34$ & $802\pm12$ & $600\pm9$ & $416\pm6$ $234\pm5$ \\ 
    $S_p [S_{\oplus}]^d$ & $4709\pm153$ & $67.5\pm2.2$ & $21.2\pm0.7$ & $4.89\pm0.16$ & $0.49\pm0.03$ \\
    $g_p^e$ [m s$^{-2}$] & $10.1\pm1.2$ & $7.1\pm0.8$ & $19.1\pm1.6$ & $19.2\pm2.3$ & - \\
    \cmidrule{1-6}
    \multicolumn{6}{c}{Common parameters} \\
    \cmidrule{1-6}
    $R_*^f [R_{\odot}]$ & \multicolumn{5}{c}{ 0.843 $\pm$ 0.005 } \\
    $M_*^f [M_{\odot}]$ & \multicolumn{5}{c}{ 0.806 $\pm$ 0.036 } \\
    $\rho_* [\rho_{\odot}]$ & \multicolumn{5}{c}{ 1.31 $\pm$ 0.05 } \\
    $u_{1, TESS}$ & \multicolumn{5}{c}{ 0.33 $\pm$ 0.08 } \\
    $u_{2, TESS}$ & \multicolumn{5}{c}{ 0.23 $\pm$ 0.09 } \\
    $u_{1, CHEOPS}$ & \multicolumn{5}{c}{ 0.46 $\pm$ 0.07 } \\
    $u_{2, CHEOPS}$ & \multicolumn{5}{c}{ 0.22 $\pm$ 0.09 } \\
    $\sigma_{HARPS-N}^g [m s^{-1}]$ & \multicolumn{5}{c}{$1.35 \pm 0.16$} \\
    $\sigma_{HIRES}^g [m s^{-1}]$ & \multicolumn{5}{c}{$2.84 \pm 0.36$} \\
    $\gamma_{HARPS-N}^h [m s^{-1}]$ & \multicolumn{5}{c}{$79699.32 \pm 0.25$} \\
    $\gamma_{HIRES}^h [m s^{-1}]$ & \multicolumn{5}{c}{$-1.27 \pm 0.42$} \\
    \bottomrule
\end{tabular}}
\begin{flushleft}
$^a$ TESS Barycentric Julian Date (BJD - 2457000). $^b$ Minimum mass in the hypothesis of a planetary origin. $^c$ Computed as $T_{eq} = T_* \left(\frac{R_{*}}{2a} \right)^{1/2}\left[(1-A_b) \right]^{1/4}$, assuming null Bond albedo ($A_b = 0$). $^d$ Stellar irradiation at the surface. $^e$Planetary surface gravity. $^f$ As determined from the stellar analysis in \citet{lacedelli2022}. $^g$ RV jitter term. $^h$ RV offset.
\end{flushleft}
\end{table*}

\section{Improvement of the ephemerides}\label{sec:ephemerides}
\begin{figure*}
    \centering
    \includegraphics[width=\textwidth]{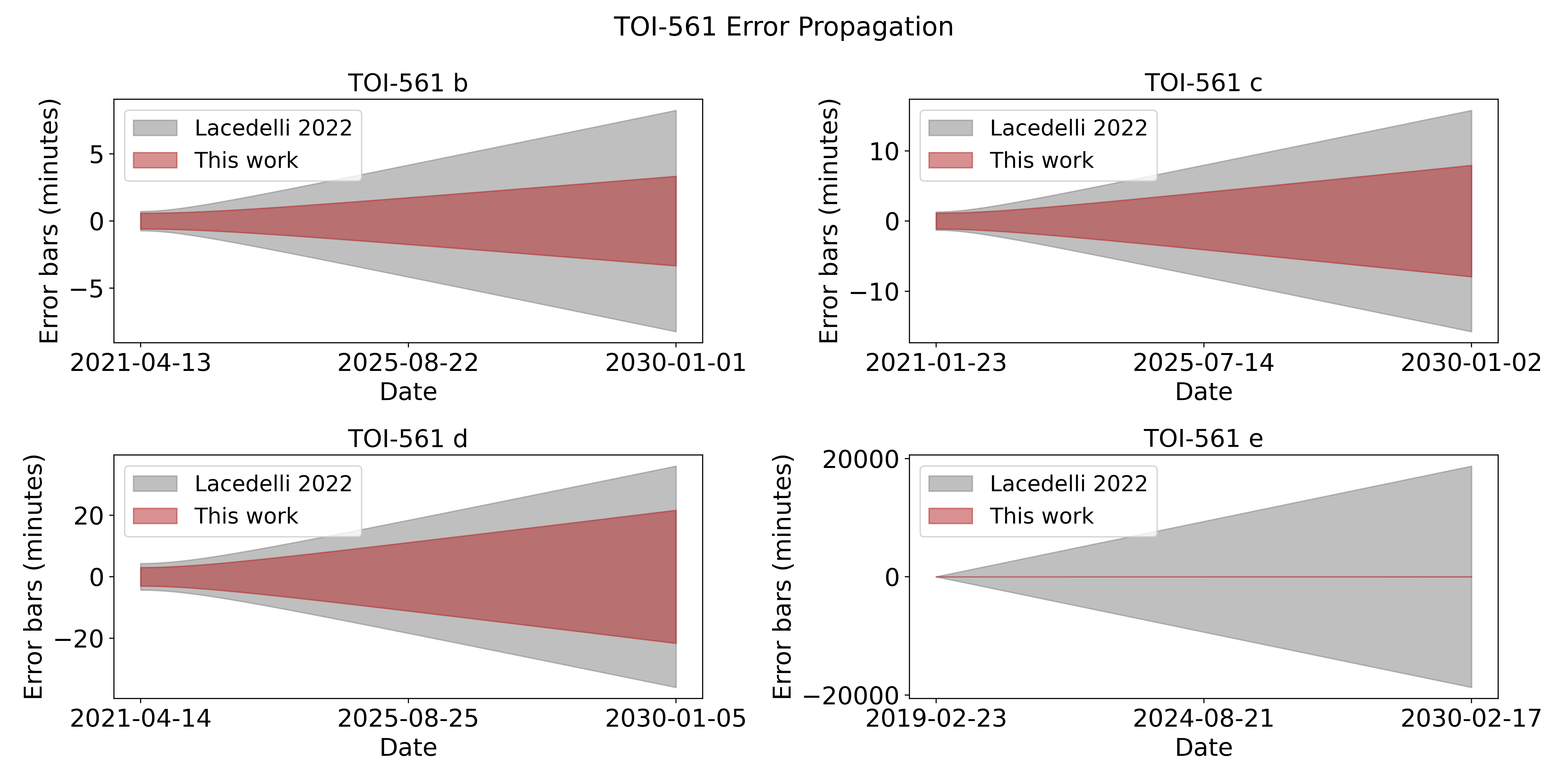}
    \includegraphics[width=0.5\textwidth]{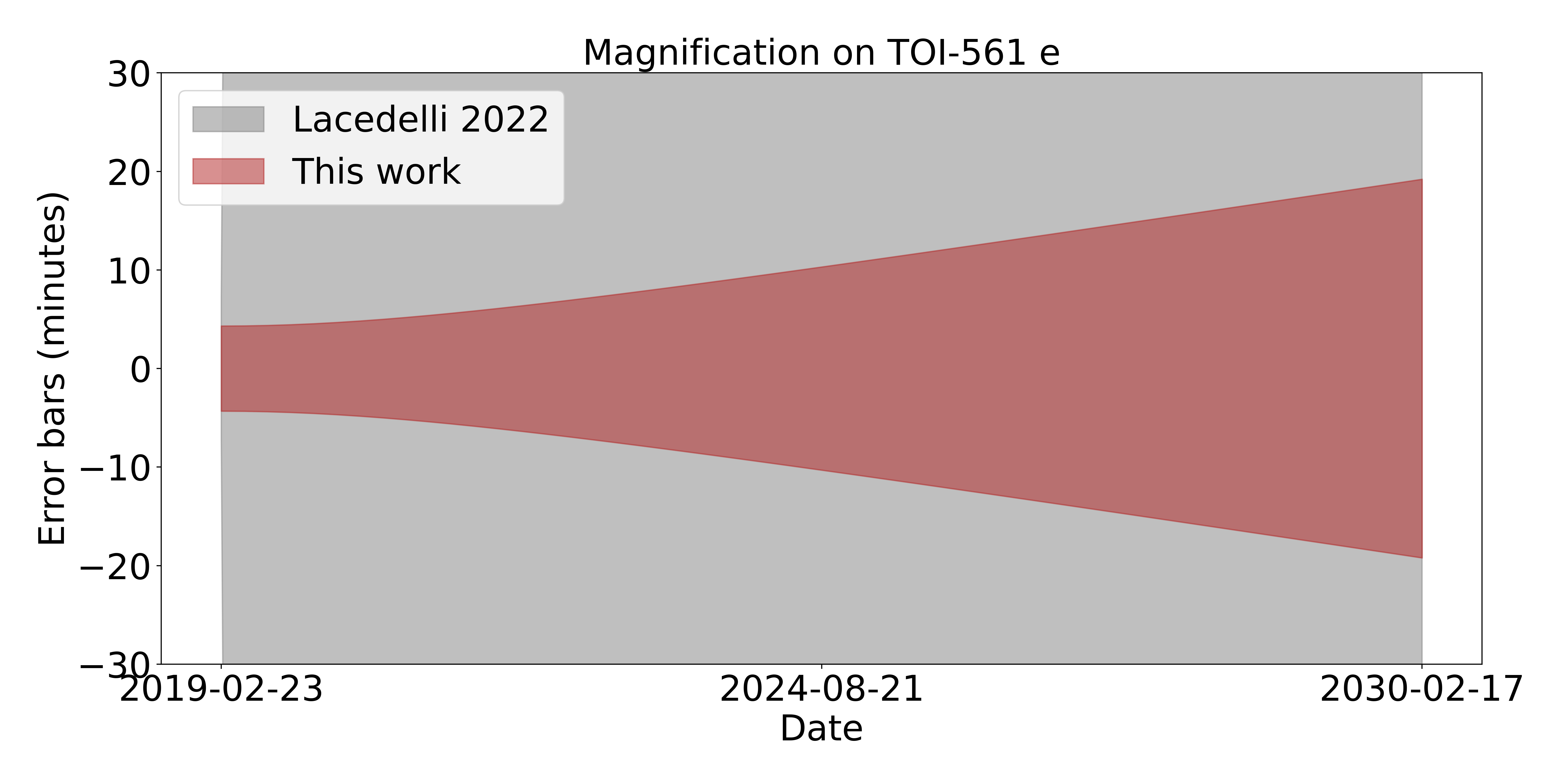}
    \caption{Upper part: Error bars on the central transit time propagated until 01 Jan 2030. Note the improvement in mid-transit time for all planets. In particular, we present for the first time plausible ephemerides of planet TOI 561 e (with an error bar on mid-transit time 974 times lower than in \citet{lacedelli2022}, at 2030). This is important for possible follow-up of this planet, e.g. for TTV and TDV measurements. Bottom part: Magnification of the planet TOI-561 e}
    \label{fig:err_prop}
\end{figure*}

Thanks to the new data set above described, we improved the transit time precision and, the ephemerides of all four transiting planets. The most remarkable achievement has been to constrain the TOI-561 e period. In Fig \ref{fig:err_prop} we show the propagated error bar for the mid-transit time of each transiting planet orbiting TOI-561. We propagated the error bars until the first available transit after the date 01 Jan 2030 and compared our results with previous ephemerides from \citet{lacedelli2022}: TOI-561 b error bar at 01-01-2030 is 4.1 minutes (using \citealp{lacedelli2022} it would be 8.2 minutes), our propagated error is 50\% smaller; TOI-561 c error bar at 01-01-2030 is 13.2 minutes; TOI-561 d error bar at 05-01-2030 is 23.6 minutes; TOI-561 e error bar at 17-02-2030 is 19.2 minutes (in \citealp{lacedelli2022} it would be 13.0 days). The new uncertainty on transit time in 2030 is 974 times (99.9\%) smaller. This last improvement is fundamental for a future follow-up observation. With new TESS and CHEOPS data, the propagation of error bars for the mid-transit times has been significantly refined.



\section{Stability analysis}\label{sec:stability}

The TOI-561 system hosts four confirmed planets (Table~\ref{tab:toi-561pars}).
In addition, radial-velocity data from HARPS-N has reported a $433 \pm 20$~day signal with a semi-amplitude of about $K=1.88 \pm 0.25$~m/s \citep{lacedelli2021}, that may belong to a fifth external companion, corresponding to a planet-$f$ with a mass around 19~$M_\oplus$.

\begin{figure*}
    \centering
	\includegraphics[width=\textwidth]{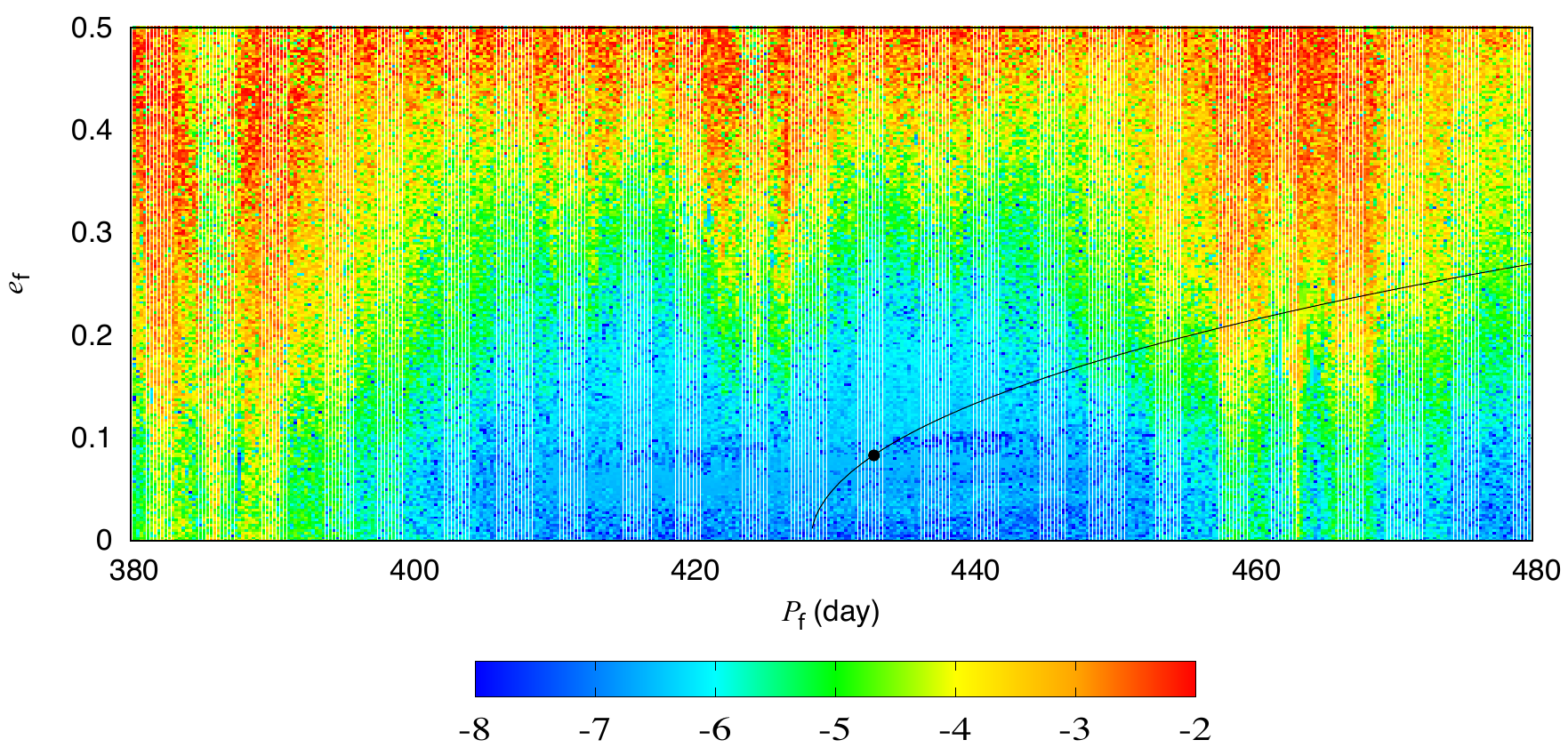}
    \caption{Stability analysis of the TOI-561 planetary system. For fixed initial conditions (Table~\ref{tab:toi-561pars}), the parameter space of the system is explored by varying the orbital period and the eccentricity of planet-$f$. The step size is $0.25$~day in the orbital period and $0.0025$ in the eccentricity. For each initial condition, the system is integrated over $10\,000$~yr, and a stability indicator is calculated, which involved a frequency analysis of the mean longitude of the inner planet. The chaotic diffusion is measured by the variation in the main frequency. Red points correspond to highly unstable orbits, while blue points correspond to orbits that are likely to be stable on Gyr timescales. The black curve corresponds to $K=1.88$~m/s and the black dot shows the values of the best fit solution (Table~\ref{tab:toi-561pars}).}
    \label{TOI-561-DA}
\end{figure*}

To check the reliability of this additional planet, we performed a stability analysis of the five-planet system in a similar way as for other exoplanetary systems \citep[eg.][]{Correia_etal_2005, Correia_etal_2010}.
In Fig.~\ref{TOI-561-DA}, we numerically explore the stability of the system on a grid of initial conditions around the best fit (Table~\ref{tab:toi-561pars}), by varying the orbital period and the eccentricity of the tentative planet-$f$, which is the most uncertain one.
The whole figure covers the $1 \sigma$ uncertainty that we have for the semi-amplitude $K$, and also $3 \sigma$ for the orbital period.
The black curve corresponds to $K=1.88$~m/s and the best-fit solution is marked with a dot.

Each initial condition was integrated for $10\,000$~yr, using the symplectic integrator SABAC4 \citep{Laskar_Robutel_2001}, with a step size of $2 \times 10^{-5} $~yr and general relativity corrections.
We then performed a frequency analysis \citep{Laskar_1990, Laskar_1993PD} of the mean longitude of planet-$f$ over two consecutive time intervals of 5000~yr, and determine the main frequency, $n$ and $n'$, respectively.
The stability is measured by $\Delta = |1-n'/n|$, which estimates the chaotic diffusion of the orbits.
The results are reported in color: orange and red represent strongly chaotic unstable trajectories, while cyan and blue give extremely stable quasi-periodic orbits on Gyr timescales.

In Fig.~\ref{TOI-561-DA}, we observe that, despite we are analysing the outermost planet, several resonances are present that disturb the system.
There is only a narrow blue stable region in the middle of the stability map, for orbital periods $P_f \approx 430 \pm 30$~day and eccentricities $e_f \lesssim 0.2$.
Interestingly, this region is where the best-fit solution from Table~\ref{tab:toi-561pars} lies.
We conclude that the TOI-561 five-planet orbital solution is plausible from a dynamical point of view, and resilient to the uncertainties in the determination of the orbital period, eccentricity, and radial-velocity semi-amplitude of planet-$f$.
Therefore, this analysis gives us an additional hint that the signal around 433~day reported in the HARPS-N data does have a planetary origin.

\section{Internal structure modelling}\label{sec:internal}

As a next step, we applied the neural network based internal structure modelling framework \texttt{plaNETic}\footnote{\url{https://github.com/joannegger/plaNETic}} \citep{Egger+2024} to the four transiting planets in the TOI-561 system, using the updated planetary parameters listed in Table~\ref{tab:toi-561pars} as well as the stellar parameters from \cite{lacedelli2022}. \texttt{plaNETic} uses neural networks to replace the computationally intensive forward model in a Bayesian accept-reject sampling algorithm used to infer the internal structure of observed planets. The neural networks used in \texttt{plaNETic} were trained on the forward model of BICEPS \citep{Haldemann+2024} and include a much more physically accurate set of equations of state as well as an envelope of fully mixed H/He and water, in contrast to the interior models previously applied to these planets \citep{lacedelli2022,patel2023}, which used a condensed water layer with a separately modelled H/He envelope.

When inferring the internal structure of observed planets, the inherent degeneracy of the problem makes the resulting posteriors dependent on the chosen priors. To account for this effect, we ran a total of six different models per planet, with two different prior options for the water content of each planet and three prior options for a planet's Si/Mg/Fe ratios in relation to the stellar abundances measured for TOI-561 \citep[see][]{lacedelli2022}. For the water priors, we chose one option compatible with a scenario where the planet formed outside the iceline and is thus expected to be water-rich, as well as an option where the planet is expected to be water-poor. For the Si/Mg/Fe priors, we applied one prior assuming the planetary abundance ratios to match the stellar ones exactly \citep[e.g.][]{thiabaud2015}, one assuming the planet to be enriched in iron compared to the star \citep[e.g.][]{adibekyan2021}, and one modelling the planet independently of the stellar Si/Mg/Fe ratios, with a uniform prior with an upper limit of 0.75 for iron. A more detailed description of these used priors and \texttt{plaNETic} in general can be found in \cite{Egger+2024}.

\begin{figure*}
    \centering
    \includegraphics[width=\textwidth]{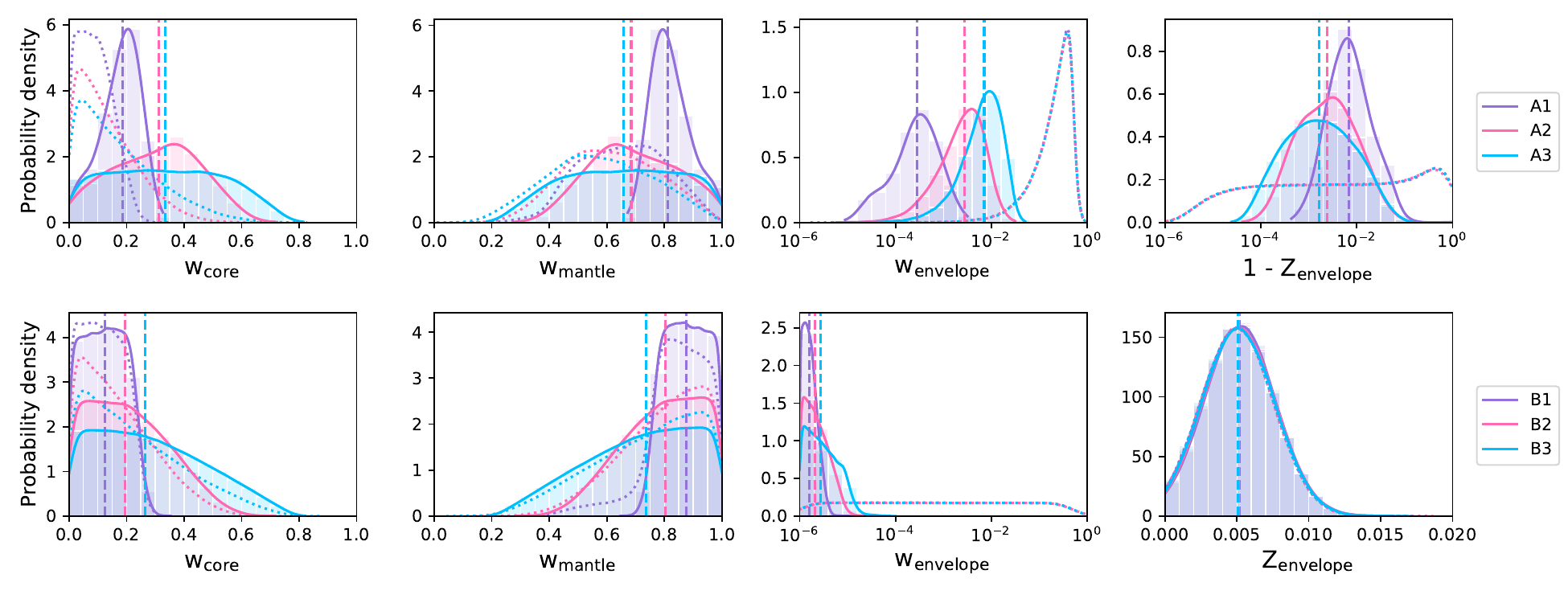}
    \caption{Results of the interior structure modelling for TOI-561~b using the \texttt{plaNETic} framework \citep{Egger+2024}. The posterior distributions shown are for the mass fractions of an inner core (w$_\mathrm{core}$), mantle (w$_\mathrm{mantle}$) and envelope (w$_\mathrm{envelope}$) in the planet as well as the mass fraction of water in the envelope (Z$_\mathrm{envelope}$). We ran models assuming a formation scenario outside the iceline, leading to a water-rich composition (top row), as well as assuming a formation scenario inside the iceline and hence a water-poor composition (bottom row). For both water prior options, we ran three different models, once assuming the planet's Si/Mg/Fe ratios to match the ones of the star exactly (purple, option 1), once assuming the planet to be more iron-enriched (pink, option 2), and once using a free uniform prior for the planet's Si/Mg/Fe ratios (blue, option 3). The dotted lines depict the chosen priors, the dashed vertical lines the median values of the posterior distributions.}
    \label{fig:internal_structure_results_b}
\end{figure*}

\begin{figure*}
    \centering
    \includegraphics[width=\textwidth]{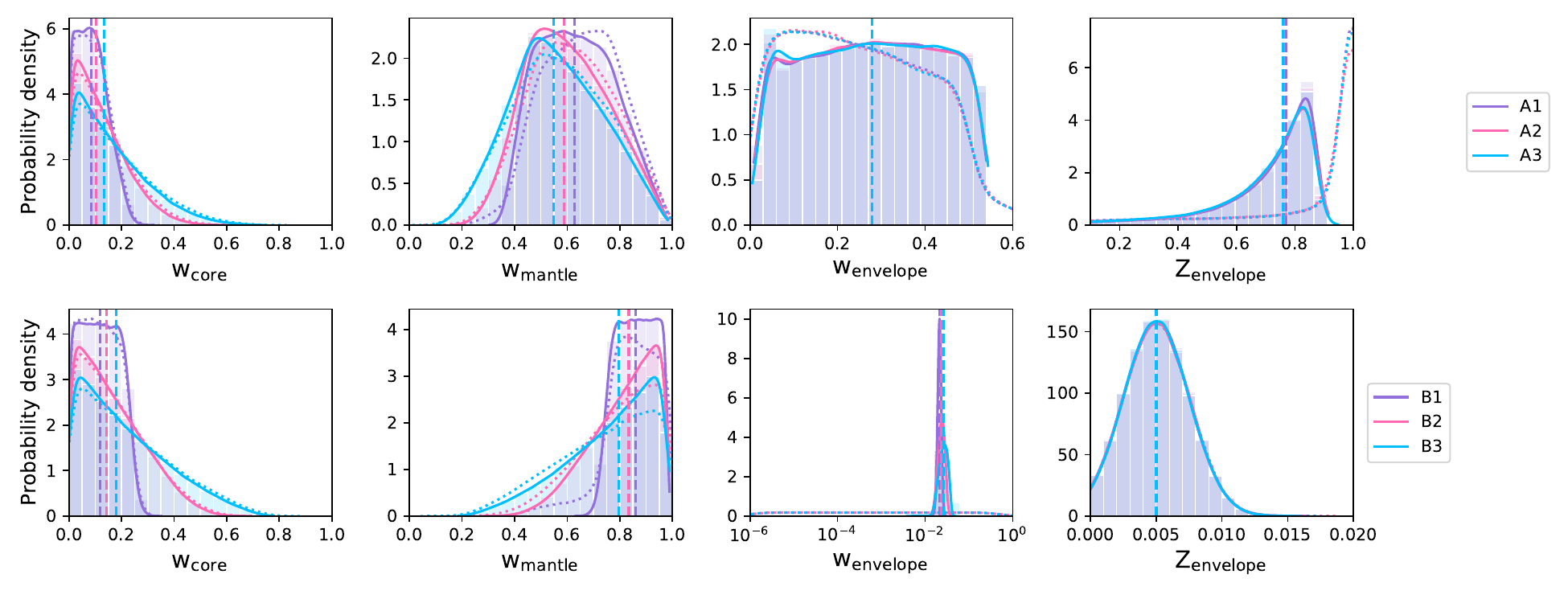}
    \caption{Same as Figure \ref{fig:internal_structure_results_b} but for TOI-561~c.}
    \label{fig:internal_structure_results_c}
\end{figure*}

\begin{figure*}
    \centering
    \includegraphics[width=\textwidth]{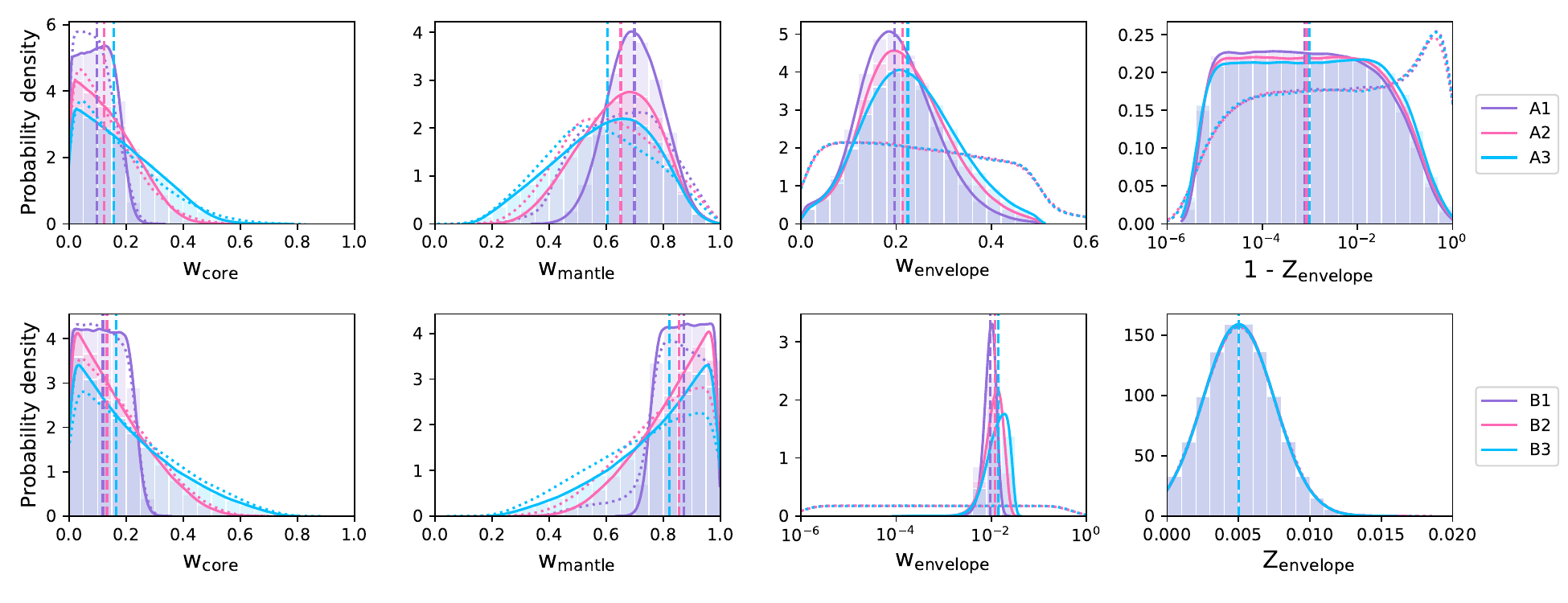}
    \caption{Same as Figure \ref{fig:internal_structure_results_b} but for TOI-561~d.}
    \label{fig:internal_structure_results_d}
\end{figure*}

\begin{figure*}
    \centering
    \includegraphics[width=\textwidth]{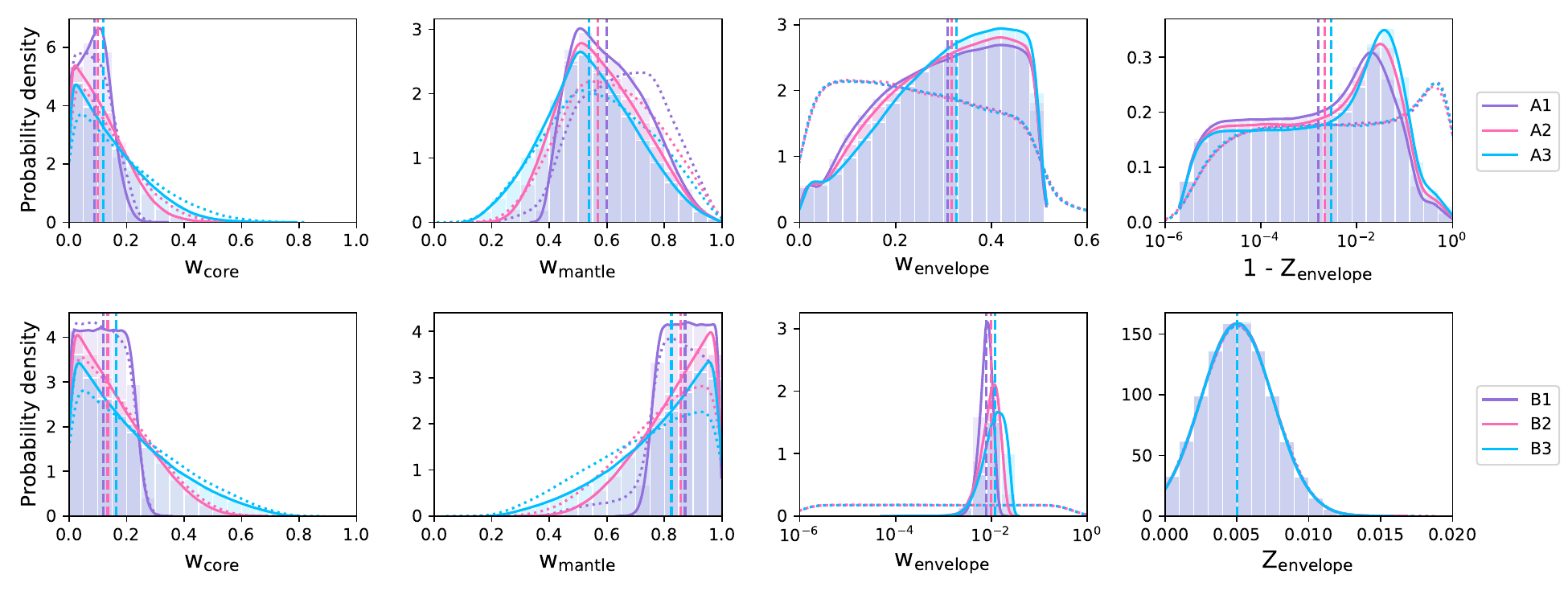}
    \caption{Same as Figure \ref{fig:internal_structure_results_b} but for TOI-561~e.}
    \label{fig:internal_structure_results_e}
\end{figure*}

The most important resulting posteriors for TOI-561~b to e are visualised in Figures~\ref{fig:internal_structure_results_b} to \ref{fig:internal_structure_results_e}. Tables~\ref{tab:internal_structure_results_b} to \ref{fig:internal_structure_results_e} summarise the median values and one sigma error intervals for all modelled internal structure parameters. For planet~b, we find that the inferred envelope mass fractions are very small for all chosen priors. If we assume a formation outside the iceline (case A), we end up with almost pure steam envelopes of <1\%. Conversely, for a formation scenario inside the iceline (case B), we infer largely H/He dominated envelopes with mass fractions of the order of $10^{-6}$, which would most likely have been evaporated. This is in agreement with the computations made in \citet{patel2023}.

If we assume that planets~c, d, and e formed inside the iceline, we expect them to host almost pure H/He envelopes with mass fractions of around 2-3\% (c) and 1-2\% (d and e). If they formed inside the iceline, we expect their envelopes to be water-rich, with water mass fractions in the envelope of around $77^{+8}_{-20}$\% for planet~c and almost pure water envelopes for planets~d and e. The mass fractions of these water-dominated envelopes are expected to be around $28^{+17}_{-18}$\% for planet~c, $21^{+10}_{-8}$\% for planet~d, and $32^{+13}_{-15}$\% for planet~e.

\section{Discussion and Conclusions}

With an age of $t_\star = 11.0_{-3.5}^{+2.8}$ Gyr \citep{lacedelli2022}, TOI-561 is one of the oldest Milky Way disk stars known to have four confirmed planets, plus an additional fifth non-transiting candidate.
TOI-561 system architecture has been controversial since its discovery. \citet{wiess2021} interpreted the transit of TOI-561 e in TESS data as the transit of a planet with $\sim16$~days period.
On the other hand, \citet{lacedelli2021} suggested that the architecture of the TOI-561 system can be explained without the planet with a 16-d period. 
Later on, \citet{lacedelli2022} found a 433 d signal in the HARPS-N RV data that could be planetary in nature. TOI-561 e, originally inferred from TESS sector 8 and HARPS-N RV data, has now been confirmed with new TESS+CHEOPS light curves, definitely fixing the architecture of the planetary system.  The masses of all planets have been measured at $\sim10\%$ precision, while the radii have been determined with $\sim2\%$ precision. 

\begin{figure}
    \centering
    \includegraphics[width=0.5\textwidth]{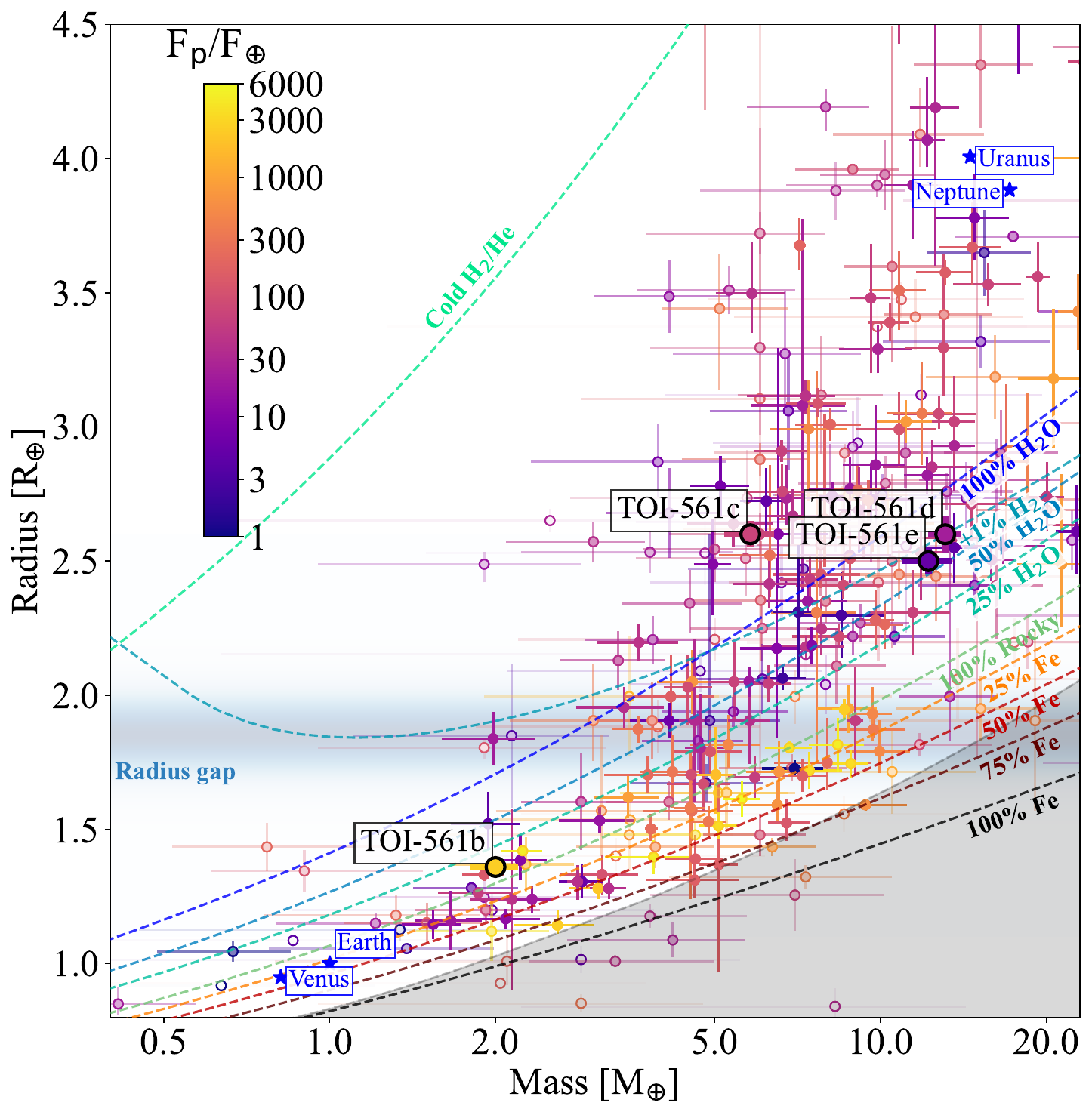}
    \caption{Mass-Radius diagram for TOI-561 system. The data are taken from the Extrasolar Planets Encyclopaedia catalog (\url{http://exoplanet.eu/catalog/}) as of May 2024. The TOI-561 planets are labeled and updated with the analysis in the current work. The theoretical mass-radius relationship as a function of the chemical composition \citep{zeng2019} are shown in dashed colored lines. The shaded grey region represents the forbidden region predicted by collisional stripping \citep{marcus2010}}
    \label{fig:MRTOI561}
\end{figure}

New CHEOPS and TESS data allowed us to estimate the density of the planets, including TOI-561 e, with unprecedented precision, and to properly place all the planets in the M-R diagram (Fig \ref{fig:MRTOI561}). 
The system includes one of the lowest density USP super-Earth TOI-561 b with a density $\rho_p = 4.3 \pm 0.6$gcm$^{-3}$.
Our analysis refined the radius and the period of the outer transiting planet TOI-561 e, which is confirmed to be a mini-Neptune with a radius of $R_p = 2.49 \pm 0.05 R_{\oplus}$ and a mass of $M_p = 12.4 \pm 1.4 M_{\oplus}$, with stellar irradiation of $4.89 \pm 0.16 S_{\oplus}$ and an equilibrium temperature of $414 \pm 3$K, assuming a bond albedo $A_B=0$.

The new observational data allowed us to improve the ephemerides of all planets, in particular we obtained a 99.9\% reduction of the uncertainty on TOI-561 e ephemerides.
TOI-561 e is one of a few good candidates for exomoon search.
With a period of about 77 days, it is above the period threshold of 60 days,
suggested by \citet{alvarado-montes2017MNRAS.471.3019A},
below which the exomoon would collide with the planet.
As described in \citet{ehrenreich2023A&A...671A.154E},
the observation aimed to search for exomoon 
should be centered on the transit of the planet,
and should span the time equivalent of the Hill radius of the planet,
because a stable exomoon is expected to lie within the planet Hill sphere
\citep{domingos2006}.
For this reason, the improvement we obtained on the linear ephemerides of planet e
has a crucial importance for the planning and scheduling of future observations.  

Finally, the stability analysis performed in this paper demonstrated the long-term stability of the outer candidate TOI-561 f, validating even further its existence.
This paper presents a definitive characterization of the exoplanet system of TOI-561, with confirmation of the five planets and improvement of ephemerides, orbital, and physical parameters, thanks to the new data from CHEOPS and TESS. Specifically, the great improvement in transit-timing precision and the fine-tuning of internal structure models underlines the relevance of TOI-561 for ongoing and future studies.


\appendix

\begin{table*}
\renewcommand{\arraystretch}{1.5}
\caption{Median and one-sigma errors for the posterior distributions of the internal structure modelling for TOI-561 b.}
\centering
\begin{tabular}{r|ccc|ccc}
\hline \hline
Water prior &              \multicolumn{3}{c|}{Formation outside iceline (water-rich)} & \multicolumn{3}{c}{Formation inside iceline (water-poor)} \\
Si/Mg/Fe prior &           Stellar (A1) &       Iron-enriched (A2) &      Free (A3) &
                           Stellar (B1) &       Iron-enriched (B2) &      Free (B3) \\
\hline
w$_\textrm{core}$ [\%] &        $18.7_{-8.2}^{+5.6}$ &    $31.3_{-18.1}^{+14.4}$ &    $33.4_{-22.0}^{+22.6}$ &
                           $12.4_{-8.3}^{+8.2}$ &    $19.6_{-13.3}^{+16.4}$ &    $26.4_{-18.1}^{+23.2}$ \\
w$_\textrm{mantle}$ [\%] &      $81.3_{-5.7}^{+8.2}$ &    $68.5_{-14.6}^{+18.0}$ &    $65.8_{-22.6}^{+21.9}$ &
                           $87.6_{-8.2}^{+8.3}$ &    $80.4_{-16.4}^{+13.3}$ &    $73.6_{-23.2}^{+18.1}$ \\
w$_\textrm{envelope}$ [\%] &    $0.03_{-0.02}^{+0.05}$ &    $0.27_{-0.20}^{+0.38}$ &    $0.71_{-0.48}^{+0.85}$ &
                           $\left(1.6_{-0.4}^{+0.8}\right)$ $10^{-4}$ &    $\left(2.1_{-0.9}^{+2.2}\right)$ $10^{-4}$ &    $\left(2.8_{-1.4}^{+4.4}\right)$ $10^{-4}$ \\
\hline
Z$_\textrm{envelope}$ [\%] &        $99.3_{-1.3}^{+0.4}$ &    $99.8_{-0.8}^{+0.2}$ &    $99.8_{-0.8}^{+0.1}$ &
                           $0.5_{-0.2}^{+0.2}$ &    $0.5_{-0.2}^{+0.2}$ &    $0.5_{-0.2}^{+0.2}$ \\
\hline
x$_\textrm{Fe,core}$ [\%] &     $90.2_{-6.2}^{+6.9}$ &    $90.6_{-6.7}^{+6.7}$ &    $90.3_{-6.3}^{+6.6}$ &
                           $90.8_{-6.6}^{+6.4}$ &    $90.7_{-6.6}^{+6.4}$ &    $90.6_{-6.6}^{+6.5}$ \\
x$_\textrm{S,core}$ [\%] &      $9.8_{-6.9}^{+6.2}$ &    $9.4_{-6.7}^{+6.7}$ &    $9.7_{-6.6}^{+6.3}$ &
                           $9.2_{-6.4}^{+6.6}$ &    $9.3_{-6.4}^{+6.6}$ &    $9.4_{-6.5}^{+6.6}$ \\
\hline
x$_\textrm{Si,mantle}$ [\%] &   $44.7_{-6.8}^{+6.5}$ &    $32.4_{-8.5}^{+9.4}$ &    $21.1_{-13.6}^{+19.8}$ &
                           $40.5_{-5.6}^{+6.0}$ &    $36.8_{-7.9}^{+7.7}$ &    $27.7_{-19.2}^{+28.0}$ \\
x$_\textrm{Mg,mantle}$ [\%] &   $46.8_{-6.9}^{+6.7}$ &    $38.6_{-9.3}^{+10.0}$ &    $30.7_{-16.0}^{+19.5}$ &
                           $47.9_{-6.2}^{+6.5}$ &    $43.1_{-9.3}^{+8.8}$ &    $34.0_{-22.1}^{+27.2}$ \\
x$_\textrm{Fe,mantle}$ [\%] &   $7.7_{-5.3}^{+7.5}$ &    $28.3_{-17.2}^{+16.5}$ &    $46.2_{-25.4}^{+16.8}$ &
                           $11.5_{-7.5}^{+6.8}$ &    $18.7_{-12.5}^{+15.7}$ &    $30.9_{-20.5}^{+23.8}$ \\
\hline
\end{tabular}
\label{tab:internal_structure_results_b}
\end{table*}
\renewcommand{\arraystretch}{1.0}

\begin{table*}
\renewcommand{\arraystretch}{1.5}
\caption{Median and one-sigma errors for the posterior distributions of the internal structure modelling for TOI-561 c.}
\centering
\begin{tabular}{r|ccc|ccc}
\hline \hline
Water prior &              \multicolumn{3}{c|}{Formation outside iceline (water-rich)} & \multicolumn{3}{c}{Formation inside iceline (water-poor)} \\
Si/Mg/Fe prior &           Stellar (A1) &       Iron-enriched (A2) &      Free (A3) &
                           Stellar (B1) &       Iron-enriched (B2) &      Free (B3) \\
\hline
w$_\textrm{core}$ [\%] &        $8.4_{-5.7}^{+6.5}$ &    $10.4_{-7.4}^{+11.8}$ &    $13.3_{-9.7}^{+16.5}$ &
                           $11.8_{-8.0}^{+8.1}$ &    $14.2_{-10.2}^{+15.5}$ &    $17.9_{-13.0}^{+21.3}$ \\
w$_\textrm{mantle}$ [\%] &      $62.9_{-14.9}^{+16.0}$ &    $58.7_{-14.7}^{+18.1}$ &    $54.9_{-16.5}^{+19.7}$ &
                           $86.1_{-8.2}^{+8.1}$ &    $83.4_{-15.8}^{+10.3}$ &    $79.5_{-21.7}^{+13.2}$ \\
w$_\textrm{envelope}$ [\%] &    $27.8_{-17.6}^{+17.1}$ &    $27.9_{-17.5}^{+17.1}$ &    $27.9_{-17.6}^{+17.1}$ &
                           $2.1_{-0.2}^{+0.2}$ &    $2.4_{-0.4}^{+0.4}$ &    $2.6_{-0.6}^{+0.7}$ \\
\hline
Z$_\textrm{envelope}$ [\%] &        $77.1_{-19.6}^{+7.8}$ &    $76.3_{-20.1}^{+8.2}$ &    $75.9_{-20.5}^{+8.2}$ &
                           $0.5_{-0.2}^{+0.2}$ &    $0.5_{-0.3}^{+0.2}$ &    $0.5_{-0.2}^{+0.2}$ \\
\hline
x$_\textrm{Fe,core}$ [\%] &     $90.3_{-6.4}^{+6.6}$ &    $90.3_{-6.4}^{+6.5}$ &    $90.3_{-6.3}^{+6.5}$ &
                           $90.3_{-6.4}^{+6.6}$ &    $90.3_{-6.4}^{+6.6}$ &    $90.3_{-6.3}^{+6.6}$ \\
x$_\textrm{S,core}$ [\%] &      $9.7_{-6.6}^{+6.4}$ &    $9.7_{-6.5}^{+6.4}$ &    $9.7_{-6.5}^{+6.3}$ &
                           $9.7_{-6.6}^{+6.4}$ &    $9.7_{-6.6}^{+6.4}$ &    $9.7_{-6.6}^{+6.3}$ \\
\hline
x$_\textrm{Si,mantle}$ [\%] &   $40.5_{-5.7}^{+6.0}$ &    $38.1_{-7.8}^{+7.6}$ &    $33.2_{-22.8}^{+29.4}$ &
                           $40.5_{-5.7}^{+6.0}$ &    $38.3_{-7.8}^{+7.5}$ &    $33.9_{-23.3}^{+29.2}$ \\
x$_\textrm{Mg,mantle}$ [\%] &   $47.9_{-6.1}^{+6.4}$ &    $44.8_{-9.1}^{+8.1}$ &    $36.2_{-24.6}^{+30.1}$ &
                           $47.9_{-6.1}^{+6.4}$ &    $45.0_{-9.0}^{+8.2}$ &    $36.4_{-24.7}^{+30.0}$ \\
x$_\textrm{Fe,mantle}$ [\%] &   $11.5_{-7.6}^{+6.7}$ &    $15.5_{-10.8}^{+15.4}$ &    $22.0_{-15.9}^{+24.0}$ &
                           $11.6_{-7.6}^{+6.7}$ &    $15.1_{-10.6}^{+15.2}$ &    $21.3_{-15.4}^{+23.6}$ \\
\hline
\end{tabular}
\label{tab:internal_structure_results_c}
\end{table*}
\renewcommand{\arraystretch}{1.0}

\begin{table*}
\renewcommand{\arraystretch}{1.5}
\caption{Median and one-sigma errors for the posterior distributions of the internal structure modelling for TOI-561 d.}
\centering
\begin{tabular}{r|ccc|ccc}
\hline \hline
Water prior &              \multicolumn{3}{c|}{Formation outside iceline (water-rich)} & \multicolumn{3}{c}{Formation inside iceline (water-poor)} \\
Si/Mg/Fe prior &           Stellar (A1) &       Iron-enriched (A2) &      Free (A3) &
                           Stellar (B1) &       Iron-enriched (B2) &      Free (B3) \\
\hline
w$_\textrm{core}$ [\%] &        $9.7_{-6.6}^{+6.5}$ &    $12.4_{-8.7}^{+12.0}$ &    $15.8_{-11.2}^{+16.4}$ &
                           $11.9_{-8.1}^{+8.2}$ &    $13.3_{-9.6}^{+15.4}$ &    $16.6_{-12.1}^{+21.4}$ \\
w$_\textrm{mantle}$ [\%] &      $69.9_{-9.5}^{+9.9}$ &    $65.1_{-15.3}^{+13.0}$ &    $60.4_{-19.8}^{+16.1}$ &
                           $87.1_{-8.2}^{+8.1}$ &    $85.5_{-15.7}^{+9.7}$ &    $82.0_{-21.8}^{+12.4}$ \\
w$_\textrm{envelope}$ [\%] &    $19.7_{-7.4}^{+8.6}$ &    $21.3_{-8.1}^{+9.7}$ &    $22.4_{-9.2}^{+10.5}$ &
                           $1.0_{-0.3}^{+0.3}$ &    $1.2_{-0.5}^{+0.5}$ &    $1.4_{-0.6}^{+0.8}$ \\
\hline
Z$_\textrm{envelope}$ [\%] &        $99.9_{-2.7}^{+0.1}$ &    $99.9_{-3.0}^{+0.1}$ &    $99.9_{-3.6}^{+0.1}$ &
                           $0.5_{-0.2}^{+0.2}$ &    $0.5_{-0.2}^{+0.2}$ &    $0.5_{-0.2}^{+0.2}$ \\
\hline
x$_\textrm{Fe,core}$ [\%] &     $90.3_{-6.4}^{+6.5}$ &    $90.4_{-6.4}^{+6.5}$ &    $90.4_{-6.4}^{+6.5}$ &
                           $90.3_{-6.4}^{+6.5}$ &    $90.4_{-6.4}^{+6.5}$ &    $90.3_{-6.4}^{+6.5}$ \\
x$_\textrm{S,core}$ [\%] &      $9.7_{-6.5}^{+6.4}$ &    $9.6_{-6.5}^{+6.4}$ &    $9.6_{-6.5}^{+6.4}$ &
                           $9.7_{-6.5}^{+6.4}$ &    $9.6_{-6.5}^{+6.4}$ &    $9.7_{-6.5}^{+6.4}$ \\
\hline
x$_\textrm{Si,mantle}$ [\%] &   $40.6_{-5.7}^{+6.0}$ &    $38.1_{-7.8}^{+7.5}$ &    $32.9_{-22.8}^{+30.1}$ &
                           $40.6_{-5.6}^{+6.0}$ &    $38.8_{-7.8}^{+7.4}$ &    $35.9_{-24.4}^{+29.4}$ \\
x$_\textrm{Mg,mantle}$ [\%] &   $47.9_{-6.2}^{+6.4}$ &    $44.7_{-8.9}^{+8.3}$ &    $37.8_{-24.8}^{+28.3}$ &
                           $47.8_{-6.1}^{+6.3}$ &    $45.5_{-9.0}^{+8.0}$ &    $36.4_{-24.8}^{+30.0}$ \\
x$_\textrm{Fe,mantle}$ [\%] &   $11.4_{-7.5}^{+6.8}$ &    $15.7_{-11.0}^{+15.0}$ &    $21.4_{-15.2}^{+22.5}$ &
                           $11.6_{-7.6}^{+6.7}$ &    $13.9_{-10.0}^{+15.2}$ &    $19.3_{-14.2}^{+23.7}$ \\
\hline
\end{tabular}
\label{tab:internal_structure_results_d}
\end{table*}
\renewcommand{\arraystretch}{1.0}

\begin{table*}
\renewcommand{\arraystretch}{1.5}
\caption{Median and one-sigma errors for the posterior distributions of the internal structure modelling for TOI-561 e.}
\centering
\begin{tabular}{r|ccc|ccc}
\hline \hline
Water prior &              \multicolumn{3}{c|}{Formation outside iceline (water-rich)} & \multicolumn{3}{c}{Formation inside iceline (water-poor)} \\
Si/Mg/Fe prior &           Stellar (A1) &       Iron-enriched (A2) &      Free (A3) &
                           Stellar (B1) &       Iron-enriched (B2) &      Free (B3) \\
\hline
w$_\textrm{core}$ [\%] &        $8.8_{-5.8}^{+5.5}$ &    $10.0_{-7.1}^{+10.6}$ &    $11.8_{-8.6}^{+14.5}$ &
                           $12.0_{-8.2}^{+8.2}$ &    $13.5_{-9.7}^{+15.3}$ &    $16.4_{-12.0}^{+21.0}$ \\
w$_\textrm{mantle}$ [\%] &      $59.9_{-11.9}^{+15.2}$ &    $56.8_{-13.2}^{+16.4}$ &    $53.7_{-15.2}^{+17.0}$ &
                           $87.2_{-8.2}^{+8.2}$ &    $85.6_{-15.6}^{+9.8}$ &    $82.4_{-21.4}^{+12.2}$ \\
w$_\textrm{envelope}$ [\%] &    $30.9_{-15.4}^{+13.0}$ &    $31.7_{-15.4}^{+12.5}$ &    $32.7_{-15.2}^{+11.8}$ &
                           $0.8_{-0.2}^{+0.2}$ &    $1.0_{-0.4}^{+0.5}$ &    $1.2_{-0.6}^{+0.7}$ \\
\hline
Z$_\textrm{envelope}$ [\%] &        $99.8_{-3.2}^{+0.2}$ &    $99.8_{-4.3}^{+0.2}$ &    $99.7_{-5.2}^{+0.3}$ &
                           $0.5_{-0.2}^{+0.2}$ &    $0.5_{-0.2}^{+0.2}$ &    $0.5_{-0.2}^{+0.2}$ \\
\hline
x$_\textrm{Fe,core}$ [\%] &     $90.1_{-6.3}^{+6.7}$ &    $90.2_{-6.3}^{+6.7}$ &    $90.3_{-6.4}^{+6.6}$ &
                           $90.3_{-6.4}^{+6.5}$ &    $90.4_{-6.4}^{+6.5}$ &    $90.3_{-6.4}^{+6.5}$ \\
x$_\textrm{S,core}$ [\%] &      $9.9_{-6.7}^{+6.3}$ &    $9.8_{-6.7}^{+6.3}$ &    $9.7_{-6.6}^{+6.4}$ &
                           $9.7_{-6.5}^{+6.4}$ &    $9.6_{-6.5}^{+6.4}$ &    $9.7_{-6.5}^{+6.4}$ \\
\hline
x$_\textrm{Si,mantle}$ [\%] &   $40.8_{-5.7}^{+6.1}$ &    $38.8_{-7.8}^{+7.3}$ &    $33.7_{-23.5}^{+30.6}$ &
                           $40.7_{-5.7}^{+6.0}$ &    $38.8_{-7.8}^{+7.3}$ &    $36.2_{-24.5}^{+29.3}$ \\
x$_\textrm{Mg,mantle}$ [\%] &   $48.2_{-6.2}^{+6.3}$ &    $45.8_{-9.2}^{+8.0}$ &    $37.1_{-25.2}^{+31.4}$ &
                           $47.8_{-6.1}^{+6.3}$ &    $45.5_{-8.9}^{+8.0}$ &    $36.6_{-24.9}^{+30.0}$ \\
x$_\textrm{Fe,mantle}$ [\%] &   $10.7_{-7.1}^{+7.1}$ &    $13.4_{-9.6}^{+15.7}$ &    $18.9_{-14.0}^{+25.9}$ &
                           $11.5_{-7.5}^{+6.7}$ &    $13.9_{-10.0}^{+15.1}$ &    $19.0_{-13.9}^{+23.3}$ \\
\hline
\end{tabular}
\label{tab:internal_structure_results_e}
\end{table*}
\renewcommand{\arraystretch}{1.0}

\section*{Acknowledgements}
CHEOPS is an ESA mission in partnership with Switzerland with important contributions to the payload and the ground segment from Austria, Belgium, France, Germany, Hungary, Italy, Portugal, Spain, Sweden, and the United Kingdom. The CHEOPS Consortium would like to gratefully acknowledge the support received by all the agencies, offices, universities, and industries involved. Their flexibility and willingness to explore new approaches were essential to the success of this mission. CHEOPS data analysed in this article will be made available in the CHEOPS mission archive (\url{https://cheops.unige.ch/archive_browser/}). 
LBo, GBr, VNa, IPa, GPi, RRa, GSc, VSi, and TZi acknowledge support from CHEOPS ASI-INAF agreement n. 2019-29-HH.0. 
TZi acknowledges NVIDIA Academic Hardware Grant Program for the use of the Titan V GPU card and the Italian MUR Departments of Excellence grant 2023-2027 “Quantum Frontiers”. 
This work has been carried out within the framework of the NCCR PlanetS supported by the Swiss National Science Foundation under grants 51NF40\_182901 and 51NF40\_205606. 
ACMC acknowledges support from the FCT, Portugal, through the CFisUC projects UIDB/04564/2020 and UIDP/04564/2020, with DOI identifiers 10.54499/UIDB/04564/2020 and 10.54499/UIDP/04564/2020, respectively. 
A.C., A.D., B.E., K.G., and J.K. acknowledge their role as ESA-appointed CHEOPS Science Team Members. 
A. S. acknowledges support from the Swiss Space Office through the ESA PRODEX program. 
S.G.S. acknowledge support from FCT through FCT contract nr. CEECIND/00826/2018 and POPH/FSE (EC). 
The Portuguese team thanks the Portuguese Space Agency for the provision of financial support in the framework of the PRODEX Programme of the European Space Agency (ESA) under contract number 4000142255. 
PM acknowledges support from STFC research grant number ST/R000638/1. 
TWi acknowledges support from the UKSA and the University of Warwick. 
YAl acknowledges support from the Swiss National Science Foundation (SNSF) under grant 200020\_192038. 
NCSa acknowledges funding by the European Union (ERC, FIERCE, 101052347). Views and opinions expressed are however those of the author(s) only and do not necessarily reflect those of the European Union or the European Research Council. Neither the European Union nor the granting authority can be held responsible for them. 
LMS gratefully acknowledges financial support from the CRT foundation under Grant No. 2018.2323 ‘Gaseous or rocky? Unveiling the nature of small worlds’. 
We acknowledge financial support from the Agencia Estatal de Investigación of the Ministerio de Ciencia e Innovación MCIN/AEI/10.13039/501100011033 and the ERDF “A way of making Europe” through projects PID2019-107061GB-C61, PID2019-107061GB-C66, PID2021-125627OB-C31, and PID2021-125627OB-C32, from the Centre of Excellence “Severo Ochoa” award to the Instituto de Astrofísica de Canarias (CEX2019-000920-S), from the Centre of Excellence “María de Maeztu” award to the Institut de Ciències de l’Espai (CEX2020-001058-M), and from the Generalitat de Catalunya/CERCA programme. 
We acknowledge financial support from the Agencia Estatal de Investigación of the Ministerio de Ciencia e Innovación MCIN/AEI/10.13039/501100011033 and the ERDF “A way of making Europe” through projects PID2019-107061GB-C61, PID2019-107061GB-C66, PID2021-125627OB-C31, and PID2021-125627OB-C32, from the Centre of Excellence “Severo Ochoa'' award to the Instituto de Astrofísica de Canarias (CEX2019-000920-S), from the Centre of Excellence “María de Maeztu” award to the Institut de Ciències de l’Espai (CEX2020-001058-M), and from the Generalitat de Catalunya/CERCA programme. 
S.C.C.B. acknowledges support from FCT through FCT contracts nr. IF/01312/2014/CP1215/CT0004. 
ABr was supported by the SNSA. 
C.B. acknowledges support from the Swiss Space Office through the ESA PRODEX program. 
ACC acknowledges support from STFC consolidated grant number ST/V000861/1, and UKSA grant number ST/X002217/1. 
P.E.C. is funded by the Austrian Science Fund (FWF) Erwin Schroedinger Fellowship, program J4595-N. 
This project was supported by the CNES. 
A.De. 
This work was supported by FCT - Funda\c{c}\~{a}o para a Ci\^{e}ncia e a Tecnologia through national funds and by FEDER through COMPETE2020 through the research grants UIDB/04434/2020, UIDP/04434/2020, 2022.06962.PTDC. 
O.D.S.D. is supported in the form of work contract (DL 57/2016/CP1364/CT0004) funded by national funds through FCT. 
B.-O. D. acknowledges support from the Swiss State Secretariat for Education, Research and Innovation (SERI) under contract number MB22.00046. 
A.C., A.D., B.E., K.G., and J.K. acknowledge their role as ESA-appointed CHEOPS Science Team Members. 
This project has received funding from the Swiss National Science Foundation for project 200021\_200726. It has also been carried out within the framework of the National Centre of Competence in Research PlanetS supported by the Swiss National Science Foundation under grant 51NF40\_205606. The authors acknowledge the financial support of the SNSF. 
MF and CMP gratefully acknowledge the support of the Swedish National Space Agency (DNR 65/19, 174/18). 
DG gratefully acknowledges financial support from the CRT foundation under Grant No. 2018.2323 “Gaseousor rocky? Unveiling the nature of small worlds”. 
M.G. is an F.R.S.-FNRS Senior Research Associate. 
MNG is the ESA CHEOPS Project Scientist and Mission Representative, and as such also responsible for the Guest Observers (GO) Programme. MNG does not relay proprietary information between the GO and Guaranteed Time Observation (GTO) Programmes, and does not decide on the definition and target selection of the GTO Programme. 
CHe acknowledges support from the European Union H2020-MSCA-ITN-2019 under Grant Agreement no. 860470 (CHAMELEON). 
KGI is the ESA CHEOPS Project Scientist and is responsible for the ESA CHEOPS Guest Observers Programme. She does not participate in, or contribute to, the definition of the Guaranteed Time Programme of the CHEOPS mission through which observations described in this paper have been taken, nor to any aspect of target selection for the programme. 
K.W.F.L. was supported by Deutsche Forschungsgemeinschaft grants RA714/14-1 within the DFG Schwerpunkt SPP 1992, Exploring the Diversity of Extrasolar Planets. 
This work was granted access to the HPC resources of MesoPSL financed by the Region Ile de France and the project Equip@Meso (reference ANR-10-EQPX-29-01) of the programme Investissements d'Avenir supervised by the Agence Nationale pour la Recherche. 
ML acknowledges support of the Swiss National Science Foundation under grant number PCEFP2\_194576. 
This work was also partially supported by a grant from the Simons Foundation (PI Queloz, grant number 327127). 
GyMSz acknowledges the support of the Hungarian National Research, Development and Innovation Office (NKFIH) grant K-125015, a PRODEX Experiment Agreement No. 4000137122, the Lend\"ulet LP2018-7/2021 grant of the Hungarian Academy of Science and the support of the city of Szombathely. 
V.V.G. is an F.R.S-FNRS Research Associate. 
JV acknowledges support from the Swiss National Science Foundation (SNSF) under grant PZ00P2\_208945. 
NAW acknowledges UKSA grant ST/R004838/1. 

\section*{Data Availability}
HARPS-N observations and data products are available through the Data \& Analysis Center for Exoplanets (DACE) at \url{https://dace.unige.ch/}. TESS data products can be accessed through the official NASA website \url{https://heasarc.gsfc.nasa.gov/docs/tess/data-access.html}. All underlying data are available either in the appendix/online supporting material or will be available via VizieR at CDS.

%
%
\bibliographystyle{mnras}
\bibliography{biblio}

\section*{Affiliations}
$^{1}$ Dipartimento di Fisica e Astronomia "Galileo Galilei", Università degli Studi di Padova, Vicolo dell'Osservatorio 3, 35122 Padova, Italy \\
$^{2}$ INAF, Osservatorio Astronomico di Padova, Vicolo dell'Osservatorio 5, 35122 Padova, Italy \\
$^{3}$ Weltraumforschung und Planetologie, Physikalisches Institut, University of Bern, Gesellschaftsstrasse 6, 3012 Bern, Switzerland \\
$^{4}$ CFisUC, Departamento de Física, Universidade de Coimbra, 3004-516 Coimbra, Portugal \\
$^{5}$ Center for Space and Habitability, University of Bern, Gesellschaftsstrasse 6, 3012 Bern, Switzerland \\
$^{6}$ Department of Astronomy, Stockholm University, AlbaNova University Center, 10691 Stockholm, Sweden \\
$^{7}$ Instituto de Astrofisica e Ciencias do Espaco, Universidade do Porto, CAUP, Rua das Estrelas, 4150-762 Porto, Portugal \\
$^{8}$ Astrophysics Group, Lennard Jones Building, Keele University, Staffordshire, ST5 5BG, United Kingdom \\
$^{9}$ Department of Physics, University of Warwick, Gibbet Hill Road, Coventry CV4 7AL, United Kingdom \\
$^{10}$ Space Research Institute, Austrian Academy of Sciences, Schmiedlstrasse 6, A-8042 Graz, Austria \\
$^{11}$ Department of Astronomy \& Astrophysics, University of Chicago, Chicago, IL 60637, USA \\
$^{12}$ Departamento de Fisica e Astronomia, Faculdade de Ciencias, Universidade do Porto, Rua do Campo Alegre, 4169-007 Porto, Portugal \\
$^{13}$ Cavendish Laboratory, JJ Thomson Avenue, Cambridge CB3 0HE, UK \\
$^{14}$ Dipartimento di Fisica, Universita degli Studi di Torino, via Pietro Giuria 1, I-10125, Torino, Italy \\
$^{15}$ Institute of Planetary Research, German Aerospace Center (DLR), Rutherfordstrasse 2, 12489 Berlin, Germany \\
$^{16}$ Observatoire astronomique de l'Université de Genève, Chemin Pegasi 51, 1290 Versoix, Switzerland \\
$^{17}$ Instituto de Astrofísica de Canarias, Vía Láctea s/n, 38200 La Laguna, Tenerife, Spain \\
$^{19}$ Departamento de Astrofísica, Universidad de La Laguna, Astrofísico Francisco Sanchez s/n, 38206 La Laguna, Tenerife, Spain \\
$^{20}$ Admatis, 5. Kandó Kálmán Street, 3534 Miskolc, Hungary \\
$^{21}$ Depto. de Astrofísica, Centro de Astrobiología (CSIC-INTA), ESAC campus, 28692 Villanueva de la Cañada (Madrid), Spain \\
$^{22}$ Centre for Exoplanet Science, SUPA School of Physics and Astronomy, University of St Andrews, North Haugh, St Andrews KY16 9SS, UK \\
$^{23}$ INAF, Osservatorio Astrofisico di Torino, Via Osservatorio, 20, I-10025 Pino Torinese To, Italy \\
$^{24}$ Centre for Mathematical Sciences, Lund University, Box 118, 221 00 Lund, Sweden \\
$^{25}$ Aix Marseille Univ, CNRS, CNES, LAM, 38 rue Frédéric Joliot-Curie, 13388 Marseille, France \\
$^{26}$ ELTE Gothard Astrophysical Observatory, 9700 Szombathely, Szent Imre h. u. 112, Hungary \\
$^{27}$ SRON Netherlands Institute for Space Research, Niels Bohrweg 4, 2333 CA Leiden, Netherlands \\
$^{28}$ Centre Vie dans l’Univers, Faculté des sciences, Université de Genève, Quai Ernest-Ansermet 30, 1211 Genève 4, Switzerland \\
$^{29}$ Leiden Observatory, University of Leiden, PO Box 9513, 2300 RA Leiden, The Netherlands \\
$^{30}$ Department of Space, Earth and Environment, Chalmers University of Technology, Onsala Space Observatory, 439 92 Onsala, Sweden \\
$^{31}$ Dipartimento di Fisica, Università degli Studi di Torino, via Pietro Giuria 1, I-10125, Torino, Italy \\
$^{32}$ National and Kapodistrian University of Athens, Department of Physics, University Campus, Zografos GR-157 84, Athens, Greece \\
$^{33}$ Astrobiology Research Unit, Université de Liège, Allée du 6 Août 19C, B-4000 Liège, Belgium \\
$^{34}$ Department of Astrophysics, University of Vienna, Türkenschanzstrasse 17, 1180 Vienna, Austria \\
$^{35}$ European Space Agency (ESA), European Space Research and Technology Centre (ESTEC), Keplerlaan 1, 2201 AZ Noordwijk, The Netherlands \\
$^{36}$ Institute for Theoretical Physics and Computational Physics, Graz University of Technology, Petersgasse 16, 8010 Graz, Austria \\
$^{37}$ Konkoly Observatory, Research Centre for Astronomy and Earth Sciences, 1121 Budapest, Konkoly Thege Miklós út 15-17, Hungary \\
$^{38}$ ELTE E\"otv\"os Lor\'and University, Institute of Physics, P\'azm\'any P\'eter s\'et\'any 1/A, 1117 Budapest, Hungary \\
$^{39}$ Lund Observatory, Division of Astrophysics, Department of Physics, Lund University, Box 118, 22100 Lund, Sweden \\
$^{40}$ IMCCE, UMR8028 CNRS, Observatoire de Paris, PSL Univ., Sorbonne Univ., 77 av. Denfert-Rochereau, 75014 Paris, France \\
$^{41}$ Institut d'astrophysique de Paris, UMR7095 CNRS, Université Pierre \& Marie Curie, 98bis blvd. Arago, 75014 Paris, France \\
$^{42}$ Dipartimento di Fisica, Università di Trento, Via Sommarive 14, 38123 Povo \\
$^{44}$ INAF, Osservatorio Astrofisico di Catania, Via S. Sofia 78, 95123 Catania, Italy \\
$^{45}$ Institute of Optical Sensor Systems, German Aerospace Center (DLR), Rutherfordstrasse 2, 12489 Berlin, Germany \\
$^{46}$ ETH Zurich, Department of Physics, Wolfgang-Pauli-Strasse 2, CH-8093 Zurich, Switzerland \\
$^{47}$ Institut fuer Geologische Wissenschaften, Freie Universitaet Berlin, Maltheserstrasse 74-100,12249 Berlin, Germany \\
$^{48}$ Institut de Ciencies de l'Espai (ICE, CSIC), Campus UAB, Can Magrans s/n, 08193 Bellaterra, Spain \\
$^{49}$ Institut d'Estudis Espacials de Catalunya (IEEC), 08860 Castelldefels (Barcelona), Spain \\
$^{50}$ Space sciences, Technologies and Astrophysics Research (STAR) Institute, Université de Liège, Allée du 6 Août 19C, 4000 Liège, Belgium \\
$^{51}$ HUN-REN-ELTE Exoplanet Research Group, Szent Imre h. u. 112., Szombathely, H-9700, Hungary \\
$^{53}$ INAF - Osservatorio Astrofisico di Catania,Oss. Astr. Catania,via S. Sofia 78,95123 Catania,Italy \\
$^{54}$ Department of Astrophysical Sciences, Princeton University, Princeton, NJ 08544, USA \\
$^{55}$ Department of Astrophysics, University of Vienna, Türkenschanzstrasse 17, 1180 Vienna, Austria

\bsp	
\label{lastpage}
\end{document}